\documentclass[fleqn,10pt]{wlscirep}
\usepackage[utf8]{inputenc}
\usepackage{booktabs,multirow}
\usepackage[T1]{fontenc}
\usepackage{bm}
\usepackage{xr}
\usepackage{cleveref}
\usepackage{soul}
\usepackage{indentfirst}

\title{A first-passage approach to diffusion-influenced reversible binding: insights into nanoscale signaling at the presynapse}

\author[1]{Maria Reva}
\author[1*]{David A. DiGregorio}
\author[2*]{Denis S. Grebenkov}

\affil[1]{Institut Pasteur, Unit of Synapse and Circuit Dynamics, CNRS UMR 3571, Paris, France}
\affil[2]{Laboratoire de Physique de la Mati\`{e}re Condens\'{e}e (UMR 7643) \\ 
CNRS -- Ecole Polytechnique, IP Paris, 91128 Palaiseau, France}

\affil[*]{senior authors}

\affil[**]{correspondence: denis.grebenkov@polytechnique.edu and david.digregorio@pasteur.fr}

\def\x{\bm{x}}
\def\kon{k_{\rm on}}
\def\koff{k_{\rm off}}
\def\P{\mathbb{P}}

\def\konU{\rm{mM}^{-1} \rm{ms}^{-1}}
\def\koffU{\rm{ms}^{-1}}
\def\DU{\mu{\rm m}^{2}{\rm ms}^{-1}}

\def\e{\cdot 10^}
\def\nm{\rm{nm}}

\def\X{{\bm{X}}}
\def\W{{\bm{W}}}

\def\t{{\mathcal T}}
\def\R{{\mathbb R}}
\def\E{{\mathbb E}}
\def\C{{\mathbb C}}
\def\ve{\varepsilon}

\keywords{diffusion-reaction system, synaptic vesicle release, first passage times}

\begin{abstract}

Synaptic transmission between neurons is governed by a cascade of stochastic reaction-diffusion events that lead to calcium-induced vesicle release of neurotransmitter. Since experimental measurements of such systems are challenging due their nanometer and sub-millisecond scale, numerical simulations remain the principal tool for studying calcium dependent synaptic vesicle fusion, despite limitations of time-consuming calculations. In this paper we develop an analytical solution to rapidly explore dynamical stochastic reaction-diffusion problems, based on first-passage times. This is the first analytical model that accounts simultaneously for relevant statistical features of calcium ion diffusion, buffering, and its binding/unbinding reaction with a vesicular sensor. In particular, unbinding kinetics are shown to have a major impact on the calcium sensor’s occupancy probability on a millisecond scale and therefore cannot be neglected. Using Monte Carlo simulations we validated our analytical solution for instantaneous calcium influx and that through voltage-gated calcium channels. Overall we present a fast and rigorous analytical tool to study simplified reaction-diffusion systems that allow a systematic exploration of the biophysical parameters at a molecular scale, while correctly accounting for the statistical nature of molecular interactions within cells, that can also serve as a building block for more general cell signaling simulators.

\end{abstract}
\begin{document}

\flushbottom
\maketitle

\thispagestyle{empty}

\section*{Introduction}

Intracellular transport of molecules is crucial for the normal function and growth of living cells \cite{alberts2008specialized}. Many intracellular signaling cascades, such as those mediated by calcium ions ($Ca^{2+}$), are generated by biochemical reactions and diffusion, which are often hard to accurately measure, particularly on the submillisecond and submicron temporal and spatial scales \cite{berridge2003calcium}. Mathematical modeling can be used to interpret and predict features of intracellular signaling that are not yet directly observable. One particularly interesting signaling process is the ability of neurons in the brain to communicate to each other by transforming electrical into chemical signals and then back to electrical signals at specialized junctions called synapses. Electrical impulses, or action potentials (APs) recruit voltage gated calcium channels (VGCC) that mediate $Ca^{2+}$ fluxes across the membrane followed by diffusion and binding to buffer molecules throughout the presynaptic terminal. The free $Ca^{2+}$ then reach and bind the target $Ca^{2+}$sensor proteins, which are tethered to neurotransmitter containing synaptic vesicles (SVs), it will trigger SV fusion with the plasma membrane and release neurotransmitter into the synaptic cleft. The released neurotransmitter molecules diffuse within the synaptic cleft to bind neurotransmitter-gated ion channels on the postsynaptic cell and then initiate an electrical signal. $Ca^{2+}$ entry and diffusion to the $Ca^{2+}$sensor are thought to occur within tens of nanometers on a  sub-millisecond timescale \cite{eggermann2012nanodomain,nakamura2015nanoscale}. Intracellular $Ca^{2+}$-binding proteins can act as buffers, which play an important role in shaping the spatial and temporal dynamics of intracellular [$Ca^{2+}$] gradients (unbound or free ions) \cite{roberts1994localization}. These dynamic [$Ca^{2+}$] gradients in turn determine the time course of the $Ca^{2+}$ occupancy of the sensor. Therefore, numerical simulations of chemical reactions and $Ca^{2+}$ diffusion have been essential for understanding the spatial-temporal dynamics of the calcium ion concentration [$Ca^{2+}$] driving synaptic vesicle fusion \cite{eggermann2012nanodomain,nakamura2015nanoscale}. 

An analytical solution of the $Ca^{2+}$ reaction-diffusion equations describing the coupling between transient $Ca^{2+}$ fluxes and the occupancy of the $Ca^{2+}$ sensor for SV fusion is not possible. Therefore, deterministic \cite{matveev2004facilitation,dittrich2013excess,nakamura2015nanoscale} and stochastic \cite{dittrich2013excess,andrews2010detailed} simulations have been workhorses to study this problem. However, both strategies are time-consuming and suffer from inaccuracies under certain parameter regimes. The finite element methods \cite{nakamura2015nanoscale} do not account for the stochastic opening of VGCCs or fluctuations in $Ca^{2+}$ flux, which should be simulated explicitly in order to accurately predict vesicle fusion probability \cite{modchang2010comparison,nakamura2015nanoscale}. This motivates the use of
Monte Carlo methods

that can generally be divided in two groups: particle based and lattice based. In the particle-based methods each particle is treated individually, making computation time prohibitively expensive when the concentrations of particles are high and/or there are numerous species ( e.g. a large number of ions). The lattice-based methods divide space into voxels and treat diffusants as concentrations rather than individual particles \cite{blackwell2006efficient,chen2017parallel}. This approach can speed up the simulations, but at the price of reduced spatial and temporal resolution. Moreover, MC techniques suffer from inaccurate simulations of statistically rare events, despite recent advances in sampling methods \cite{donovan2016unbiased}. In this context, analytical  solutions of simplified models can provide new insights and intuition  into complex systems, as well as much faster simulation speeds.
In particular, analytical solutions allow for a rapid exploration of a vast parameter space to reveal relevant spatial and temporal scales of specific parameter combinations.

The classical example of such an approach is the linearized buffer approximation (LBA \cite{naraghi1997linearized}) that yields an approximate analytical solution of a diffusion-reaction problem. This method allows one to compute the concentration of calcium ions in a chosen location without numerical integration. This can be extended to multiple buffers and has provided important intuition about their spatial-temporal impact on intracellular $Ca^{2+}$, and potential effect on the probability of SV fusion. However, this approach can only be applied in the steady state conditions, thus not suitable for the brief and transient $Ca^{2+}$  influx driven by APs \cite{nakamura2018variations}.  Another recent multi-scale approach is based on the narrow escape problem \cite{holcman2014,metzler2014,grebenkov2017} of searching for a hidden target by a single calcium ion. An analytical solution of this problem was found and coupled with a Markovian jump process that models buffering and calcium influx \cite{guerrier2016hybrid}. In spite of its advantages, this hybrid method does not account for the $Ca^{2+}$sensor's binding and unbinding kinetics, which is of crucial importance for the vesicle release dynamics, as shown below (Figure 1). In turn, recent first-passage approaches have been used to account for the finite backward rate constant of binding for multiple particles \cite{grebenkov2017first,lawley2019first}, but do not consider competing binding partners for diffusants, which is necessary for biological realism and is the advantage of LBA. Finally, as all cells are known to have endogenous $Ca^{2+}$ buffers, a simulation approach that can account for diffusion and reaction with a target and competitors, is essential to model calcium-dependent SV fusion.

Here we propose a probabilistic diffusion-influenced reversible calcium binding model that overcomes the aforementioned deficiencies by considering the forward and backward binding rate constants of the $Ca^{2+}$sensor, as well as competing binding partners (fixed endogenous (EFB) and mobile buffers). This novel analytical model simulates a point source  $Ca^{2+}$ entry, reaction with buffer, diffusion and binding to a $Ca^{2+}$-binding protein that mediates SV fusion.  The solution allows us to study the effect of binding reaction rate constants on the occupancy probability of the sensor by $Ca^{2+}$ at all temporal scales with much lower computational costs than of any existing numerical alternatives (e.g. for Fig. \ref{fig:fig5}B. it took 96 hours to simulate black curve with MC methods given 250 CPUs, while less than a minute was necessary to produce its counterpart with the analytical method at a laptop with 1.7 GHz Intel Core i5). We confirm the necessity of taking into account the unbinding kinetics in the simulations of vesicle release probabilities. We also demonstrate the validity of the analytical solution by Monte Carlo simulations and study the effects of the sensor's kinetics and geometrical properties of the synapse on the probability of the single site occupancy. Moreover, an extension to multiple calcium ions and its limitations are discussed. To our knowledge, this is the first analytical solution for a stochastic reaction-diffusion problem that accounts simultaneously for target binding/unbinding kinetics in the presence of competing buffer species, and  accurately predicts target occupancies following stochastic influx from ion channels, as compared to particle-based Monte Carlo simulations. We also show that a cooperative, multiple independent binding site release sensor can also be implemented analytically. This approach is therefore applicable to a wide range of biochemical processes within cells that operate via diffusion-influenced cooperative or non-cooperative reactions.

\section*{Results}

\subsection*{Impact of unbinding kinetics on the vesicle release probability and time course}

\begin{figure}[t!]
\centering
\includegraphics[width=1\linewidth]{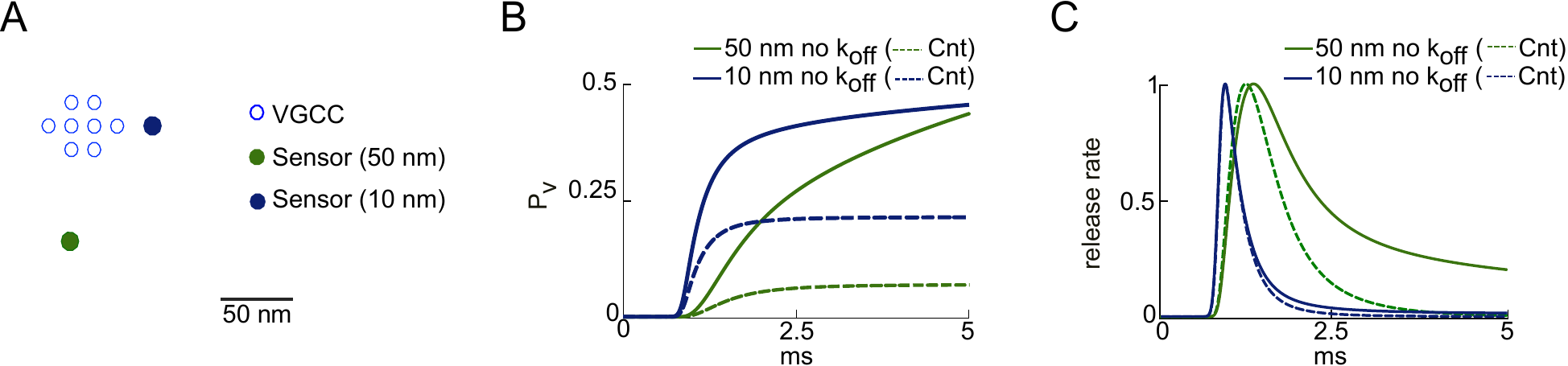}
\caption{\label{fig:fig1} \textbf{A.} Diagram of the active zone arrangement, with a cluster of VGCC (hollow circles) and two positions of SV sensors at 10 and 50 nm. \textbf{B.} Vesicle release probability for two scenarios: control (with unbinding reaction,"Cnt", dashed lines) and without unbinding reactions in the sensor kinetics (solid lines), and two CDs: 10~nm (blue) and 50~nm (green). C. Release rates (corresponding to panel B), normalized to peak amplitude.}
\end{figure}

Reversible first-order chemical reactions are described by forward ($\kon$) and backward ($\koff$) rate constants. However, in the case of $Ca^{2+}$ diffusion and binding to a $Ca^{2+}$ sensor for SV fusion, it has been argued that the first passage to the target is the dominant physical process influencing the probability of SV fusion over time, and thus approximations without $\koff$ might be sufficient. We tested the importance of $Ca^{2+}$ sensor $\koff$ for AP-evoked SV fusion by solving reaction-diffusion equations by a finite-elements method (see Methods section). Spatio-temporal profiles of free [$Ca^{2+}$]  were simulated for sensor distances of 10 and 50 nm from the perimeter of a VGCC cluster (perimeter model \cite{nakamura2015nanoscale,rebola2019distinct}, see Fig. \ref{fig:fig1}A)  in the presence of $Ca^{2+}$ buffers (ATP or endogenous fixed buffers).

We modeled the probability of SV fusion within the five binding site kinetic model of the $Ca^{2+}$ sensor \cite{wang2008synaptic} (see Methods section), and compared to a model in which $\koff$s were set to zero. For  sensor-to-channel distances (coupling distance, CD) as short as 10 nm, the time course of SV release within the first millisecond is hardly different with and without a $\koff$ (Fig. \ref{fig:fig1}C, blue lines), while the release probability is increased by 2.4 times (Fig. \ref{fig:fig1}B, blue lines). However, in the case of 50 nm CD (which is physiological at some synapses \cite{vyleta2014loose,rebola2019distinct}), setting the $\koff$ to zero increased the vesicle release probability by 7-fold (Fig. \ref{fig:fig1}B, green lines), and increased the half-width of the time course of fusion probability by 61\%  (Fig. \ref{fig:fig1}C, green lines). These simulations show that for the shortest CDs, first passage time models that only consider $\kon$ could qualitatively reproduce the time evolution in sensor occupancy, but not the final probability of SV fusion. However, for longer CDs both the time course and fusion probability were altered in the absence of unbinding. Thus, a reversible $Ca^{2+}$ binding reaction (finite $\koff$) must be considered for such simulations, particularly since the estimated coupling distances range from 10 nm to 100 nm across synapses \cite{rebola2019distinct}. 
Moreover, by solving analytically the governing reaction-diffusion equations,
we would provide an efficient framework for studying and modeling the dynamics
of molecular diffusion and binding to a partner. In the following sections, we investigate analytical solutions that describe, specifically, $Ca^{2+}$ diffusion and consider explicitly the binding and unbinding kinetics of the sensor.

\subsection*{Analytical model of $Ca^{2+}$ reaction-diffusion}

\begin{figure}[t!]
\centering
\includegraphics[width=1\linewidth]{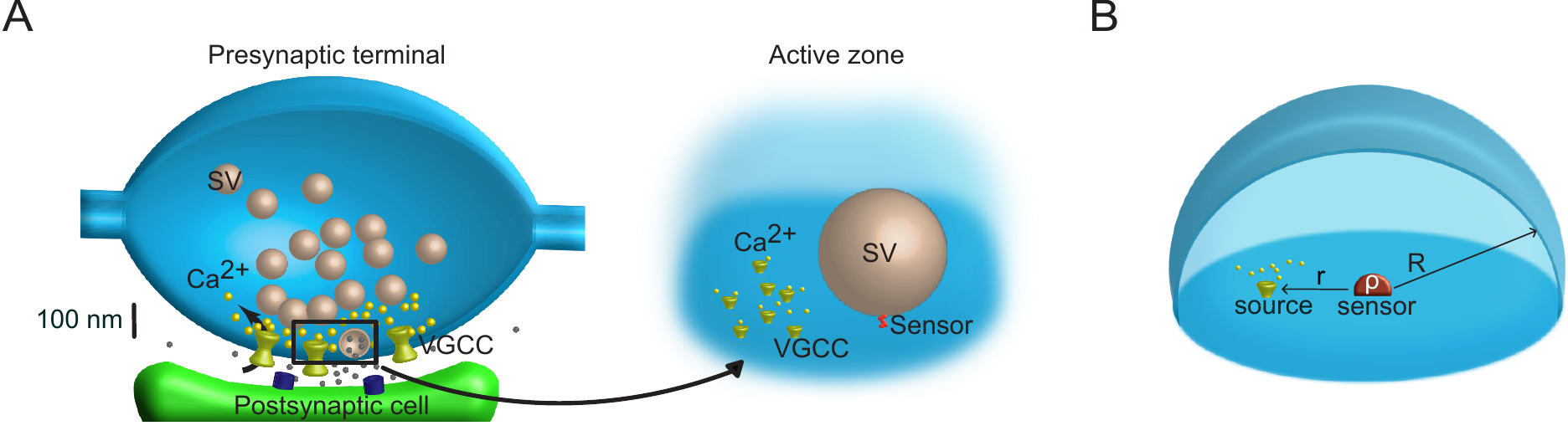}
\caption{\label{fig:chem_syn}\textbf{A.} Schematic diagram of an axonal bouton (presynaptic terminal) containing a release site (active zone). Inset: Idealized active zone scheme showing VGCC clusters and their tens of nanometers proximity to the $Ca^{2+}$ target sensor for SV fusion. \textbf{B.} Geometric representation of the simulation domain. The simulation compartment is a reflecting half sphere of radius $R$, target is a partially absorbing half sphere of radius $\rho$, the point source of $Ca^{2+}$ entry is located on the membrane (horizontal surface) at the coupling distance (CD), $r-\rho$, from the sensor (target).}
\end{figure}

 Understanding the lifetime of calcium ions on single binding sites of a SV fusion sensor and the influence of competing buffers is essential for studying the nanoscale signaling driving neurotransmitter release at synapses. As a first approximation of this process, we developed an analytical model of $Ca^{2+}$ diffusion based on a modified first passage time process, in which the target $Ca^{2+}$ sensor binding occupancy is modeled as a first order reaction with reversible kinetics (both $\kon$ and $\koff$) . This probabilistic diffusion-influenced reversible calcium binding model is described in detail in the Methods section (see also Fig. \ref{fig:chem_syn}). In brief, we placed a single $Ca^{2+}$ sensor, with the values of $\kon$ and $\koff$ taken from models predicting experimental data \cite{wang2008synaptic}, at the center of the circular surface of a half sphere.  We assumed an unlimited binding capacity of the sensor that permits each $Ca^{2+}$ binding event to occur independently.
The hindering effect of the synaptic vesicle was not considered, since it was shown not to influence sensor occupancy\cite{nakamura2018variations}. The dynamics of $Ca^{2+}$ ions is described as switching diffusion between free and buffer-bound
states\cite{yin2010,grebenkov2019}.  
In summary our model has the following parameters: the size of the sensor $\rho$, the distance between origin of the simulation domain and calcium channel $r$, the radius of the simulation domain $R$, $\kon$ and $\koff$ of the $Ca^{2+}$ sensor, the exchange
rates $k_{0i}$ and $k_{i0}$ (product of concentration and forward rate constant of the buffer) for binding/unbinding to $i$-th buffer,
and diffusion coefficients of free $Ca^{2+}$ ($D_0$)
and those bound to buffers ($D_i$ - diffusion coefficient of $i$-th buffer), see Table \ref{tab:param}.  

In order to derive a set of equations describing the solution, we took a two-step approach. First,  we found the probability distribution of the first-passage time of a $Ca^{2+}$ ion to a simplified (single binding site) sensor in the presence of competing binding partners for a single $Ca^{2+}$ entering the bouton. Second a renewal technique \cite{klafter2011first}
allowed us to incorporate unbinding kinetics on the sensor and to relate the distribution of the first-binding time to the occupancy probability, $P(t,r)$, for a single $Ca^{2+}$ ion started at some position $r$, to be on the sensor at time $t$ (see Methods, Eq. (\ref{eq:Pt}) - (\ref{eq:alpha_M0})). While our
first-passage approach is applicable to any number of co-existing
buffers, we focused on two cases of no buffer and single buffer, for
which explicit analytical formulas for $P(t,r)$ were provided. We explored the accuracy of these formulas and their assumptions using Monte Carlo
simulations (see Methods). Finally, we extended the analytical solution to account for multiple binding sites of the SV fusion sensor and for multiple $Ca^{2+}$ ions that can be released simultaneously or progressively through a single or multiple VGCCs. By solving analytically the governing reaction-diffusion equations, we provide an efficient framework for studying and modeling the dynamics of $Ca^{2+}$ diffusion and binding a molecular target, in particular AP driven $Ca^{2+}$-mediated SV fusion. 

\begin{table}[!t]
\centering
\begin{tabular}{| l | l | c | c | l |}
\hline
\textbf{Parameter} & \textbf{Notation} & \textbf{Value} & \textbf{Unit} &\textbf{ Reference} \\ \hline
\multicolumn{5}{|c|}{\textbf{Geometrical parameters}}  \\ \hline
Simulation domain radius & $R$ & 300 & $\nm$ & adapted from \cite{shahrezaei2004consequences} \\ \hline
Sensor's radius  & $\rho$ & 5 & $\nm$ & adapted from \cite{aracc2006close} \\ \hline
Coupling distance (CD) & $r - \rho$ & 15 & $\nm$ & \\ \hline
\multicolumn{5}{|c|}{}   \\    \hline
Calcium diffusion coefficient & $D_0$  & 0.22 & $\DU$ &  \cite{allbritton1992range} \\ \hline
%

\multicolumn{5}{|c|}{\textbf{$Ca^{2+}$ sensor}}  \\ \hline
Forward rate constant & $\kon$ & $5\cdot 127$ & $\konU$ &  \\   \cline{1-3}
Backward rate constant   & $\koff$ & 15.7 & $\koffU$ & \cite{wang2008synaptic} \\ \hline 
%
\multicolumn{5}{|c|}{\textbf{EFB  ($i=1$)}} \\  \hline
Diffusion coefficient & $D_1$ & 0 & $\DU$ & \\ \hline
Backward rate constant & $k_{10}$ & 10 & $\koffU$ & \\ \cline{1-4}
Forward rate constant& $k_{\rm{on},1}$ & 100 & $\konU$ & \cite{xu1997kinetic}  \\ \hline 
Total concentration & $c_1$ & 4 & mM & \cite{nakamura2015nanoscale} \\ \hline
Binding rate & $k_{01}$ & 400 & $\koffU$ &  \\ \hline
%
\multicolumn{5}{|c|}{\textbf{ATP buffer ($i=2$)}} \\ \hline
Diffusion coefficient & $D_2$ & 0.2 & $\DU$ & \\ \cline{1-4} 
Reverse rate constant & $k_{20}$ & 10 & $\koffU$ & \cite{naraghi1997linearized} \\  \cline{1-4}  
Forward rate constant & $k_{\rm{on},2}$ & 100 & $\konU$ &   \\ \hline
Total concentration & $c_2$ & 0.2 & mM & \cite{nakamura2015nanoscale} \\ \hline
Binding rate & $k_{02}$ & 20 & $\koffU$ &  \\ \hline
%
\multicolumn{5}{|c|}{\textbf{EGTA buffer ($i=3$)}} \\ \hline
Diffusion coefficient & $D_3$ & 0.22 & $\DU$ & \cite{naraghi1997linearized} \\ \hline
Backward rate constant & $k_{30}$ & 0.000735 & $\koffU$ & \cite{nagerl2000binding} \\ \cline{1-4} 
Forward rate constant & $k_{\rm{on},3}$ & 10.5 & $\konU$ &   \\  \hline
Total concentration & $c_3$ & 10 & mM & \cite{nakamura2015nanoscale} \\ \hline
Binding rate & $k_{03}$ & 105 & $\koffU$ &  \\ \hline
\end{tabular} 
\caption{\label{tab:param}Biophysical parameters of our diffusion-reaction model of the synaptic vesicle fusion sensor occupancy.}

\end{table}

\subsection*{Single ion occupancy probability for a single $Ca^{2+}$ binding site of the SV fusion sensor}

\begin{figure}[t!]
\centering
\includegraphics[width=1\linewidth]{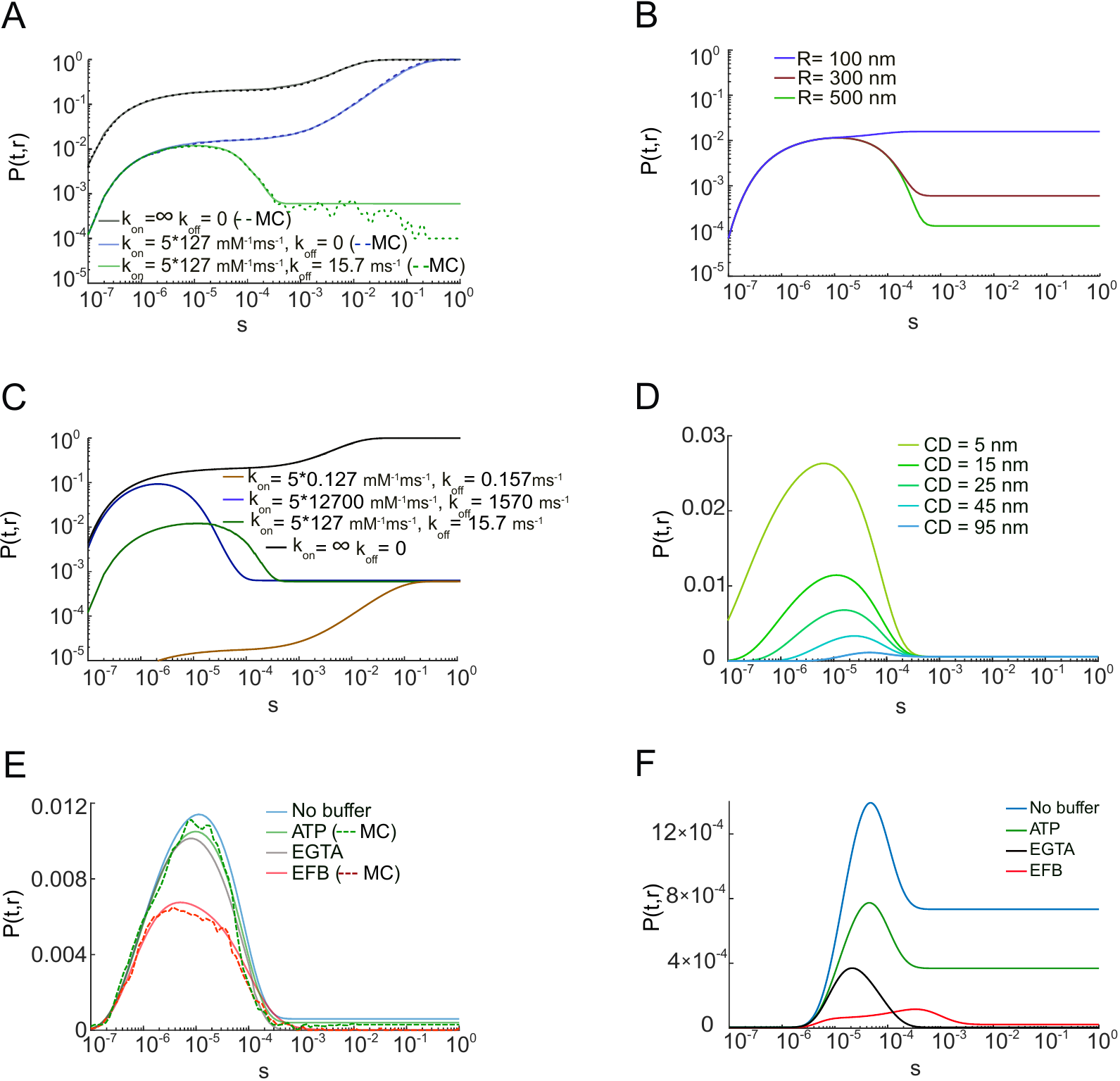}
\caption{\label{fig:fig3} The occupancy probability $P(t,r)$. \textbf{A.} $P(t,r)$ computed using analytical solution (solid lines) and MC simulations (dashed lines) for the case of instant binding (black lines); in the presence of binding kinetics (blue lines); in the presence of both referent binding and unbinding kinetics (green lines). \textbf{B.} Analytically computed $P(t,r)$ for different sizes of the domain ($R$): 100 nm (blue), 300 nm (red) and 500 nm (green). \textbf{C\textbf{}.} $P(t,r)$ computed using analytical solution for instant binding (black line), fast reaction rate constants (blue line), referent (green line) and slow reaction rate constants (brown line). \textbf{D.} Analytically computed $P(t,r)$ for CDs, $r-\rho$, varying from 5 to 95 nm.  \textbf{E.} $P(t,r)$ computed using analytical solution (solid lines) and MC simulations (dashed lines) in the absence (blue) or presence of one of the following buffers: ATP (green), EGTA (gray) and EFB (red), at the CD of 15 nm. \textbf{F.} Similar to E, but at the CD of 45 nm.}
\end{figure}

Using our analytical solution, it was possible to calculate the  occupancy probability of a single binding site of the SV sensor by a single calcium ion, $P(t,r)$,
across seven orders of magnitude in time scales, from submicroseconds to seconds, using different model parameters. For the idealized case of 
instantaneous binding and no unbinding ($\kon = \infty $, $\koff = 0$), any $Ca^{2+}$ ion that hits the sensor remains bound forever. As a consequence, $P(t,r)$ is equal to the cumulative distribution function of the first passage time to the sensor. As expected, this probability monotonically increases with time and approaches $1$ after one second (Fig. \ref{fig:fig3}A, black solid line), consistent with a pure diffusion-limited reaction. The analytical solution is in excellent agreement with MC simulations, using the same model parameters (Fig. \ref{fig:fig3}A, black dashed line). When using a finite forward rate constant ($\kon = 5\cdot 127~\konU$, $\koff = 0$), the $P(t,r)$ was reduced  (Fig. \ref{fig:fig3}A, blue solid line), in perfect agreement with MC simulations (Fig. \ref{fig:fig3}A, blue dashed line). The use of both finite forward and backward rate constants generated a biphasic occupancy curve: the $P(t,r)$ increased to a maximum value and followed by a decrease to a steady state value, as expected physiologically. The rising phase of this curve matched that of the curve when  $\koff$ was set to zero (Fig. \ref{fig:fig3}A, green solid line), thereby delineating the time scale where only ion binding is dominant.  MC simulations reproduced well the analytical solutions (Fig. \ref{fig:fig3}A, green dashed line), despite the inherent fluctuations due to a limited number of MC trials.

When increasing the size of the bouton (simulation volume), $R$=500~nm, the peak of $P(t,r)$ was not altered, but the steady-state $P(t,r)$ was decreased ($1\e{-3}$ for $R =300~\nm$, $3\e{-4}$ for $R = 500~\nm$), consistent with alteration in the steady-state, volume-averaged [$Ca^{2+}$] (Fig. \ref{fig:fig3}B).  For smaller bouton sizes ($R$=100~nm), the peak of $P(t,r)$ was increased ($2\e{-2}$). The rising phases, however,
were identical for all tested radii, suggesting that diffusion determines the initial time course of SV fusion, provided that the bouton volume is much larger than the CD. 

The probability $P(t,r)$ was then computed for different pairs of forward and backward rate constants, each chosen such that equilibrium
dissociation constant remains constant ($K_D= \kon/\koff \approx 40~ \textrm{mM}^{-1}$). In the three considered cases of fast ($\kon =
5\cdot12700~\konU$, $\koff = 1570~ \koffU$), the reference ($\kon =
5\cdot127~\konU$, $\koff = 15.7~ \koffU$), and slow
($\kon = 5\cdot0.127~\konU$, $\koff = 0.157~\koffU$) rate constants,
the $P(t,r)$ reach the same equilibrium (Fig.
\ref{fig:fig3}C). However, within the first few hundreds of microseconds,
faster rate constants enable a rapid capturing of $Ca^{2+}$ as well as faster unbinding, thus generating biphastic $P(t,r)$ that is larger, briefer and earlier  (Fig. \ref{fig:fig3}C, blue line). Interestingly 
even with a forward rate-constant of greater than $10^{10}$~M$^{-1}$~s$^{-1}$, the diffusion-limited case is not matched on the sub-microsecond time scale (Fig. \ref{fig:fig3}C, black line), due to the fast $\koff$. Thus, the $\koff$ can strongly influence target occupancy and the regime of diffusion-limited binding.

The distance between $Ca^{2+}$ sources (VGCCs) and the $Ca^{2+}$ sensor of SVs can vary across synapse types, influencing the kinetics and probability of SV fusion \cite{eggermann2012nanodomain}.  We, therefore, examined the effect of varying this VGCC-SV coupling distance in the range between 5 and 95 nm on $P(t,r)$ (Fig. \ref{fig:fig3}D). The peak of $P(t,r)$ decreased from 0.027 at 5 nm to 0.001 at 95 nm. As expected from finite elements simulations \cite{bucurenciu2008nanodomain,nakamura2015nanoscale}, 
longer CDs resulted in smaller peak fusion probabilities and longer times to peak fusion rates (6.1~$\mu$s at 5 nm to 47.5~$\mu$s at 95 nm).

Finally, we explored the effect of various $Ca^{2+}$ buffers that critically shape the spatio-temporal profile of intracellular $Ca^{2+}$. We explored the effect of ATP, a naturally occurring low-affinity, fast and mobile endogenous calcium buffer \cite{naraghi1997linearized}, non-specific low-affinity endogenous fixed buffers (EFB) \cite{xu1997kinetic}, and the mobile exogenous buffer EGTA (see Table \ref{tab:param}). EGTA is a well-characterized buffer, with slow forward $Ca^{2+}$ binding rate constant, that has been used to infer VGCC-SV CDs through competition with the $Ca^{2+}$ sensor, thus producing an observed inhibition of synaptic transmission proportional to the CD \cite{eggermann2012nanodomain, nakamura2018variations}. Because of its slow $\kon$, large concentrations of EGTA are needed to interfere with the sensor for SV fusion (larger than 1 mM and up to 10 mM \cite{nakamura2018variations}). For particle-based simulations this can be computationally prohibitive.
The effect of all three buffers have been studied extensively, and thus have well-characterized binding rate constants\cite{nakamura2015nanoscale}, see Table \ref{tab:param}.
Using our analytical approach we could rapidly calculate $P(t,r)$ for a CD of 15 nm in the presence of either 0.2 mM ATP, or 10 mM EGTA. Both buffers only slightly decreased the peak amplitude of $P(t,r)$ from 0.012 to 0.01 and shifted the time of its peak from 10~$\mu$s to 8.5~$\mu$s. On the other hand, high concentration of EFB (4 mM) had
more prominent effect, decreasing the peak probability of being bound to 
$7\e{-3}$ and shifting its time to 4~$\mu$s (Fig. \ref{fig:fig3}E). These results are consistent with the lack of effect of ATP being due to its low concentration, the lack of effect of EGTA being due to its slow forward rate constant.  At a CD of 45 nm,
the $P(t,r)$ peak was decreased from
$1\e{-3}$ to $3\e{-4}$ (EGTA), $4\e{-5}$ (EFB) and $6\e{-3}$ (ATP);
the time of peak was shifted from 4.7~$\mu$s to 2.1~$\mu$s (EGTA), 9
$\mu$s (EFB) and 4.5~$\mu$s (ATP) (Fig.
\ref{fig:fig3}F). The steady-state $Ca^{2+}$ occupancy is dramatically reduced by the large concentration of the high affinity buffer, EGTA. These differential affects of EGTA on the peak occupancy for CD of 15 and 45 nm, as well as on the steady-state occupancy, are very similar to previous analytical \cite{naraghi1997linearized}, finite elements simulations \cite{bucurenciu2008nanodomain,nakamura2015nanoscale}, and MC simulations \cite{rebola2019distinct}, due largely to the slow forward rate constant. The analytical solution was verified with MC
simulations for ATP and EFBs (Fig. \ref{fig:fig3}E), but not for EGTA as the large number of molecules associated with 10 mM EGTA was too time-consuming for MC simulations.

In fact, it was not possible to verify the analytical solution on Fig.\ref{fig:fig3}F with our MC method, due the inability of the MC simulations to capture such rare events, even when the number of trials was increased to 50 000.
These results illustrate the advantage  of the analytical approach to provide an intuitive understanding of stochastic reaction and diffusion across a wide range timescales and for large numbers of molecules, conditions that are prohibitive when using particle-based simulators.

\subsection*{Temporal regime in which reaction-diffusion models must consider reversible binding with its target}

\begin{figure}[t!]
\centering
\includegraphics[width=1\linewidth]{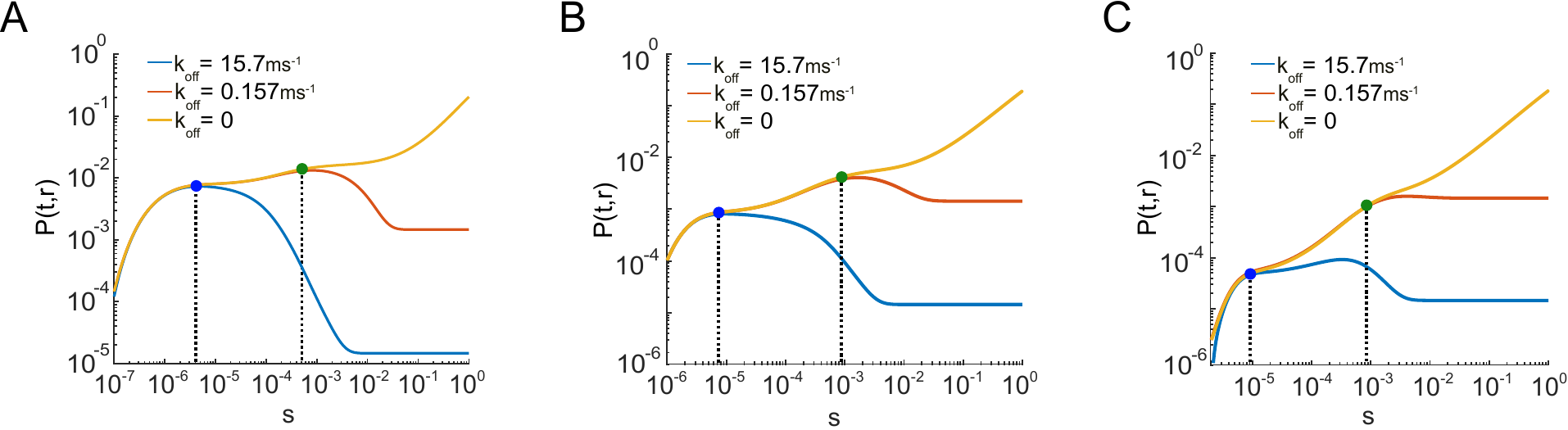}
\caption{\label{fig:fig4}  Influence of backward rate constant on $P(t,r)$. \textbf{A.} $P(t,r)$ computed using analytical solution with fixed $\kon = 5\cdot127~\konU$, and the following unbinding rates: $\koff = 0$ (yellow line), $\koff = 0.157~\koffU$ (red line) and $\koff =15.7~\koffU$ (blue line), for the CD of 15 nm. The departure between $P(t,r)$s is depicted by dots and vertical dashed lines. Blue dot: $t_c \simeq 2.1~\mu$s, green dot: $t_c \simeq 510~\mu$s.  \textbf{B\textbf{}.} Similar to A, for CD of 45 nm. Blue dot: $t_c \simeq 6.5~\mu$s, green dot: $t_c \simeq 840~\mu$s. \textbf{C.} Similar to A, for CD of 95 nm. Blue dot: $t_c \simeq 9.9~\mu$s, green dot: $t_c \simeq 980~\mu$s.} 
\end{figure}

Equipped with an analytical solution, we reexamined the importance of $\koff$ in dictating $P(r,t)$. Figure \ref{fig:fig4} shows $P(r,t)$ calculated for different $\koff$s and for different CDs: 15 nm (Fig. \ref{fig:fig4}A), 45 nm (Fig. \ref{fig:fig4}B) and 95 nm (Fig. \ref{fig:fig4}C). The high temporal resolution of the simulations show that there is a characteristic time window $(0,t_c)$ at which the sensor occupancy is independent of the $\koff$. The upper limit $t_c$ of this characteristic time window was defined as the time point when two $P(t,r)$ curves for different unbinding kinetics start to deviate (blue and green dots, Fig. \ref{fig:fig4}). This characteristic time increases as $\koff$ decreases. For CDs less than 50 nm, the characteristic time window was less than 10 microseconds for physiological rate constants (Fig. \ref{fig:fig4}A,B). The simplified first passage time approach confirms finite element simulations (Fig. \ref{fig:fig1}) showing that backward rate constants in the physiological range can influence $Ca^{2+}$ sensor's occupancy for physiological source to target distances, and therefore must be modeled explicitly for accurate predictions of $Ca^{2+}$-dependent SV fusion.

\subsection*{Sensor occupancy probability for multiple calcium ions}

\begin{figure}[t!]
\centering
\includegraphics[width=1\linewidth]{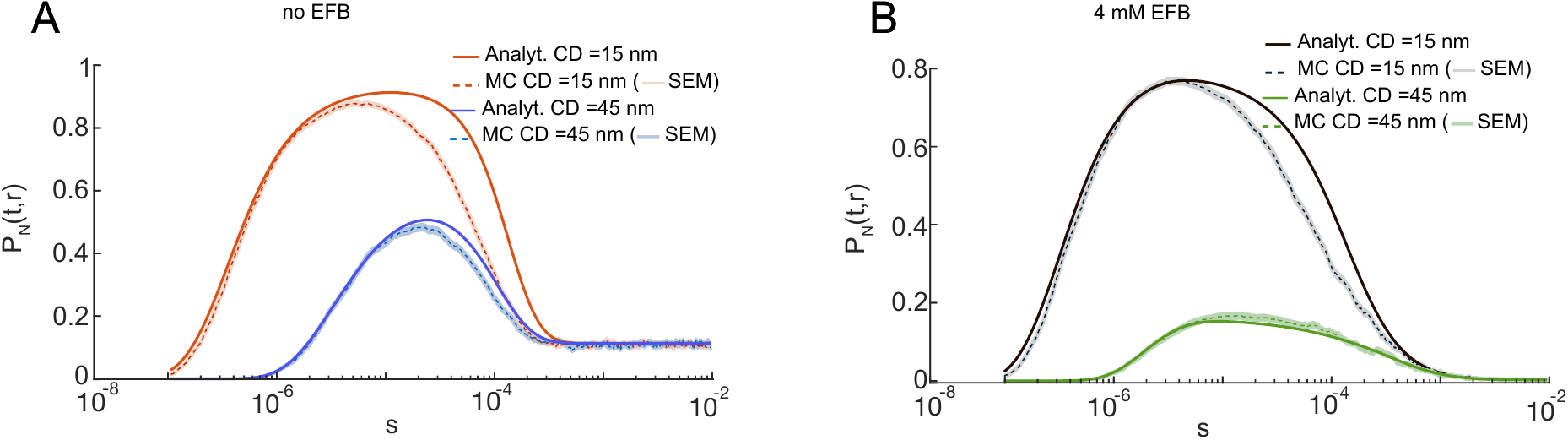}
\caption{\label{fig:fig5} Occupancy probability, $P_N(t,r)$, for an instantaneous influx of many $Ca^{2+}$ ions ($N = 200$), computed using analytical method (solid lines) and MC simulations (dashed line; SEM: shaded region). \textbf{A.} In the absence of fixed buffer, for CD of 15 nm (red lines) and CD of 45nm (blue lines). \textbf{B.} Similarly in the presence of EFB (4 mM), CD of 15 nm (black lines) and CD of 45 nm (green lines).}
\end{figure}

Thus far we considered the case when only a single $Ca^{2+}$ enters the presynaptic volume. We next explore performance of our analytical solution for $Ca^{2+}$ fluxes. During action potential-induced opening of a single channel we estimate approximately 200 $Ca^{2+}$ ions enter at a single VGCCs driving release over a Gaussian-like time course (half-width ~0.3 ms) \cite{rebola2019distinct, nakamura2018variations}. Knowing $P(r,t)$ for the single $Ca^{2+}$ ion, we used Eq. (\ref{eq:PN}) to approximate the probability $P_N(r,t)$ that at least one  $Ca^{2+}$ is bound to the sensor at time $t$ following an instantaneous influx of $N$ $Ca^{2+}$ ions (see Methods).
For an instantaneous flux of 200 ions, $P_N(t,r)$ was shown to increase to nearly $1$ for a CD of 15 nm, in contrast to the low probabilities (<0.01) in the single ion case. A similar peak was estimated using MC simulations, but the time course of  analytical $P_N(t,r)$ was broader than that computed using MC simulations
(Fig. \ref{fig:fig5}A, dashed line; shaded region is standard error of the mean (SEM)). This difference between the analytical solution and the MC simulation was smaller for the
longer CD of 45 nm (Fig. \ref{fig:fig5}A, blue lines). This discrepancy identifies a shortcoming of the analytical approximation for multiple $Ca^{2+}$ ions, which does not account for binding exclusion when the target experiences another ion while already bound. In other words,
the discrepancy between the MC and analytical
approximation can be attributed to the saturation of the single
binding site, which was not taken into account in the analytical
solution for the $Ca^{2+}$ influx.
Since the probability that two ions might interact with the target is decreased in the presence of the competing buffer molecules, we
tested whether the presence of physiological concentrations of EFBs influences the difference between the analytical
approximation and MC simulations. Indeed, the presence of EFB
decreases $P_N(t,r)$ for both CDs (Fig. \ref{fig:fig5}B), as well as the discrepancy between
analytical and simulated curves (Fig. \ref{fig:fig5}B, black lines). The error in the time course estimate was still present for shorter CDs, but for the longer CD, the two curves are indistinguishable (Fig.
\ref{fig:fig5}B, green lines). The better accuracy in the presence of EFB can be attributed to lower binding probabilities experienced by the target sensor, which is consistent with lower [$Ca^{2+}$].

\begin{figure}[t!]
\centering
\includegraphics[width=1\linewidth]{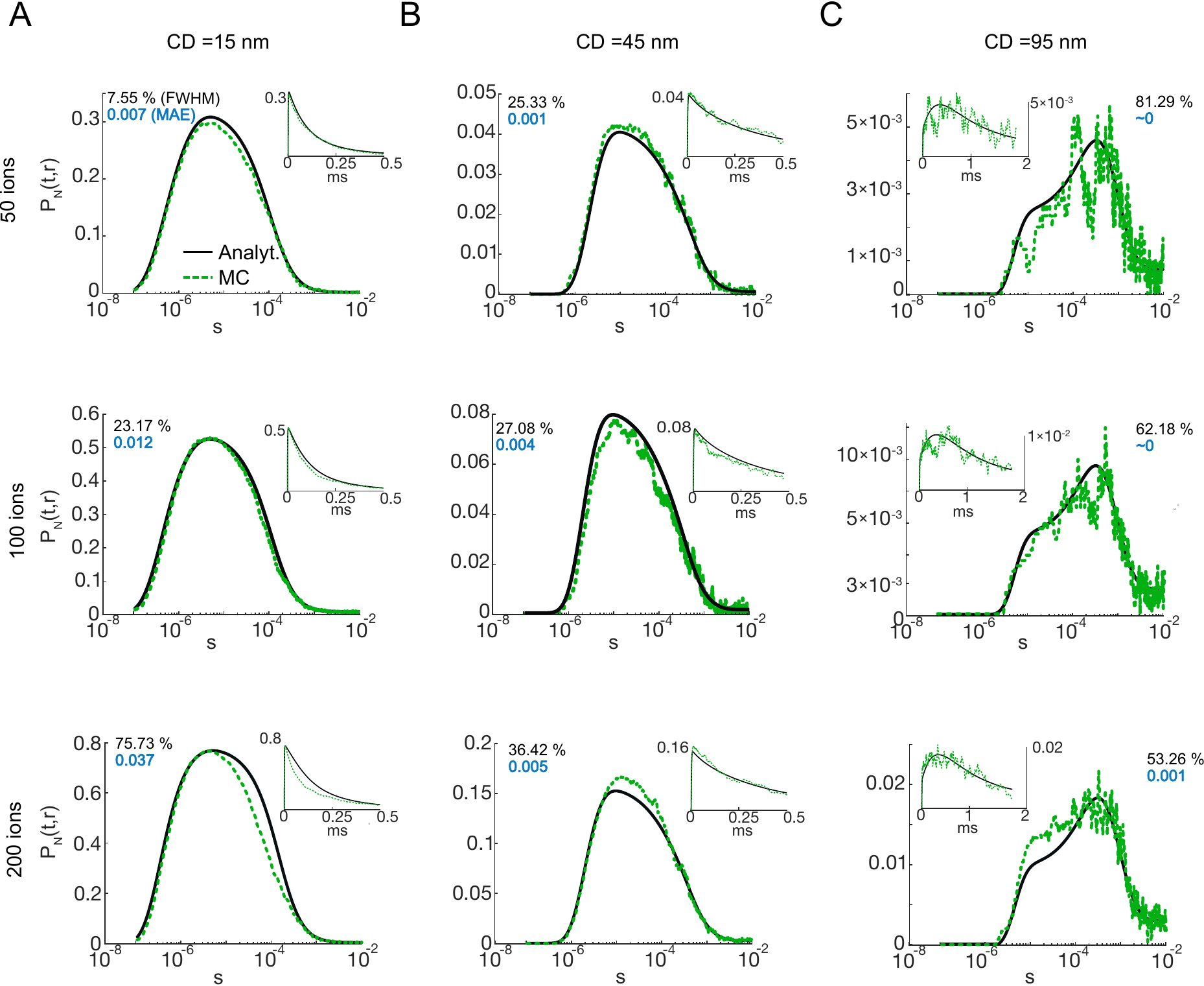}
\caption{\label{fig:fig6} $P_N(t,r)$ for various numbers of ions and differing coupling distances (CD). $P_N(t,r)$ for 50 (1st row), 100 (2nd row) and 200 (3rd row) simultaneously released ions for CD of 15 nm (\textbf{A}), 45 nm (\textbf{B}) and 95 nm (\textbf{C}). Main plots represent semi-log scale, while linear scale plots are on insets. Black and green lines show respectively analytical and MC results. The black and blue inset text on each plot represent FWHM error and MAE
between analytical and MC results correspondingly.All these computations were realized in the presence of EFB (4 nM).} 
\end{figure}

Because both the presence of a competing $Ca^{2+}$ buffer and increasing the distance to the target would be expected to reduce the occupancy probability, we next explored how the number of released $Ca^{2+}$ ions and unbinding kinetics influenced the discrepancy between MC and analytical $P_N(t,r)$ solutions. The strength of applied instantaneous $Ca^{2+}$ influx was varied (50, 100, and 200 ions) and the difference between MC and analytical curves was quantified by the Mean Absolute difference (MAE) and full width at half maximum (FWHM) error. We saw that with decreasing number of released ions, the dissimilarity decreases for all values of CDs, reflected in the values of MAE and FWHM error (Fig. \ref{fig:fig6}). However, for long CDs we notice a decrease in the FWHM error with increasing number of released ions (from 81$\%$ for 50 ions to 53$\%$ for 200 ions) (Fig. \ref{fig:fig6}C), this is due to reduced trial variability in MC simulations arising from a higher $P_N(t,r)$.
Moreover, altering $P_N(t,r)$ by adjusting $\koff$ (slow ($0.157~\koffU$) and fast ($1570~\koffU$)) was also consistent with the primary source of error being due to high occupancy (see also Figs. 1 and 2 in the Supplementary Information (SI)).

In summary, our systematic study suggests that the analytical solution can be used to compute the occupancy probability 
provided the peak sensor occupancy is less than 50\%.

\subsection*{Analytical solution for computing sensor occupancy in response to $Ca^{2+}$ fluxes generated by stochastic VGCCs}

\begin{figure}[t!]
\centering
\includegraphics[width=1\linewidth]{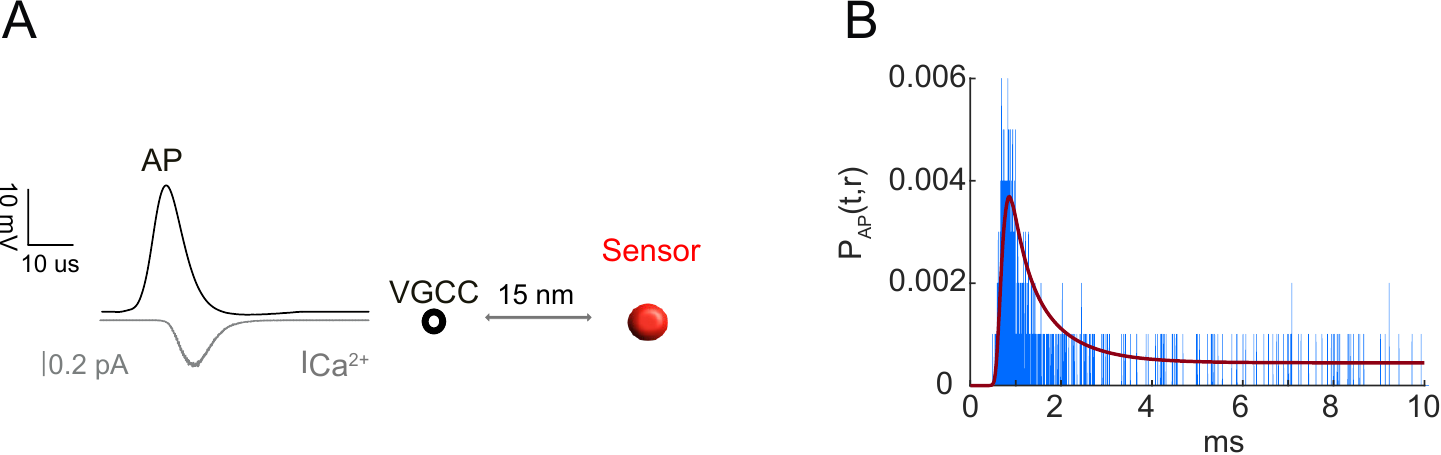}
\caption{\label{fig:fig7} 
$Ca^{2+}$ entry through a stochastic VGCC. \textbf{A.} 
Opening of the single VGCC is driven by an AP (black line) triggering $Ca^{2+}$ influx (gray line). Released $Ca^{2+}$ ions diffuse towards a sensor at distance of 15~nm. \textbf{B.} The $P_{AP}(t,r)$ computed using analytical approximation (red) and MC (blue).}
\end{figure}

Thus far we considered instantaneous entry of $Ca^{2+}$ ions, however it is known that presynaptic $Ca^{2+}$ fluxes arise from a temporally distributed opening of VGCCs during an AP, lasting hundreds of microseconds (Fig. \ref{fig:fig7}A). Moreover, it is also known that accurate estimates of the occupancy probability must consider the stochastic nature of VGCC opening, particularly as compared to deterministic approximations of mean open channel probability \cite{modchang2010comparison, nakamura2015nanoscale}.  Here, we studied the analytical solution, for $Ca^{2+}$ fluxes generated from stochastic opening of VGCCs ($P_{AP}(t,r)$; see Methods). VGGC openings and associated $Ca^{2+}$ fluxes were obtained from MC simulations. The VGCC model was constrained by experimental estimates of single channel open probability, single channel conductance and duration of the current \cite{rebola2019distinct} (see Methods and Section V of the SI). For each $Ca^{2+}$ ion entry time the $P(t,r)$s were calculated, the corresponding occupancy probability for at least one ion was found for each trial, and the occupancy probability $P_{AP}(t,r)$ was then averaged over 1000 trials (see Eq. (\ref{eq3}) in Methods). 

For a single VGCC located 15 nm from the sensor (Fig. \ref{fig:fig7}A), in the presence of EFB, the calculated $P_{AP}(t,r)$ was similar to that from MC
simulations (Fig. \ref{fig:fig7}B, based on 1000 MC trials).
The spike-like character of the simulated curve is the consequence of a finite number of trials. In fact, in the current physiological setting, the number of bound calcium ions for each trial is zero for most of time, except for a few short periods when it switches to 1 or, exceptionally, 2 bound ions. As the average duration of such periods is $1/\koff \approx 0.06$~ms, they look as narrow spikes on Fig. \ref{fig:fig7}B showing the time range between 0 and 10 ms. Moreover, as the binding times are random, the binding periods from different trials typically do not overlap and thus produce multiple individual spikes, except near the maximum of $P_{AP}(t,r)$, where they can superimpose upon the averaging. To eliminate such a discrete character, one would need to increase the number of trials considerably (up to $10^5$ or even $10^6$), which is computationally expensive. In contrast, the theoretical curve, obtained with only 1000 trials, is smooth because the averaging over infinitely many realizations is intrinsically incorporated into the notion of probability.

The peak occupancy is two orders of magnitude smaller than that from the earlier calculation for an instantaneous entry of 200 ions (Fig. \ref{fig:fig6}A), suggesting that the errors due to multiple $Ca^{2+}$ binding are minimal under physiological conditions where the ionic flux occurs over hundreds of microseconds. The extension to multiple channels is trivial using the principle of superposition \cite{naraghi1997linearized}, provided the total sensor occupancy remains less than 50\%. Thus our analytical method is capable of describing how a simplified SV fusion sensor could be driven by stochastic channel openings.

\subsection*{The occupancy probability for binding at least $n$ ions}

Seminal experiments at the frog neuromuscular synapse showed a nonlinear relationship between extracellular $[Ca^{2+}]$ and neurotransmitter release, that could be described by a Hill coefficient of ~4 (Ref. \cite{dodge1967co}). More recent evidence suggested that this nonlinearity could be due, in part, to the multi-site occupancy of the sensor protein, synaptotagmin-1, for $Ca^{2+}$  \cite{chapman2018ca2+}. Measurements of single AP-evoked SV fusion are well explained by a 5-state release model \cite{wang2008synaptic,bollmann2000calcium,schneggenburger2000intracellular} (see Methods). However, a recent mammalian SV fusion model indicated that steep Hill coefficients of SV fusion could also result from the independent binding of calcium ions to multiple single sites \cite{dittrich2013excess}, suggesting that cooperativity between binding sites is not required to model the nonlinear relationship between intracellular $[Ca^{2+}]$ and release. To keep our model analytically tractable, we considered the latter sensor model and derived an analytical solution for the occupancy probability, $P_{N,n}(t,r)$, that at least $n$ calcium ions are bound to the sensor at time $t$, given that $N$ calcium ions were released simultaneously at time $0$ (see Eq. (\ref{eq:PNn}) of Methods). If the kinetic rate constants of binding sites were identical and independent, then the $P_{N,n}(t,r)$ would be identical to the occupancy probability of at least $n$ binding sites of the sensor at time $t$. 

\begin{figure}[t!]
\centering
\includegraphics[width=1\linewidth]{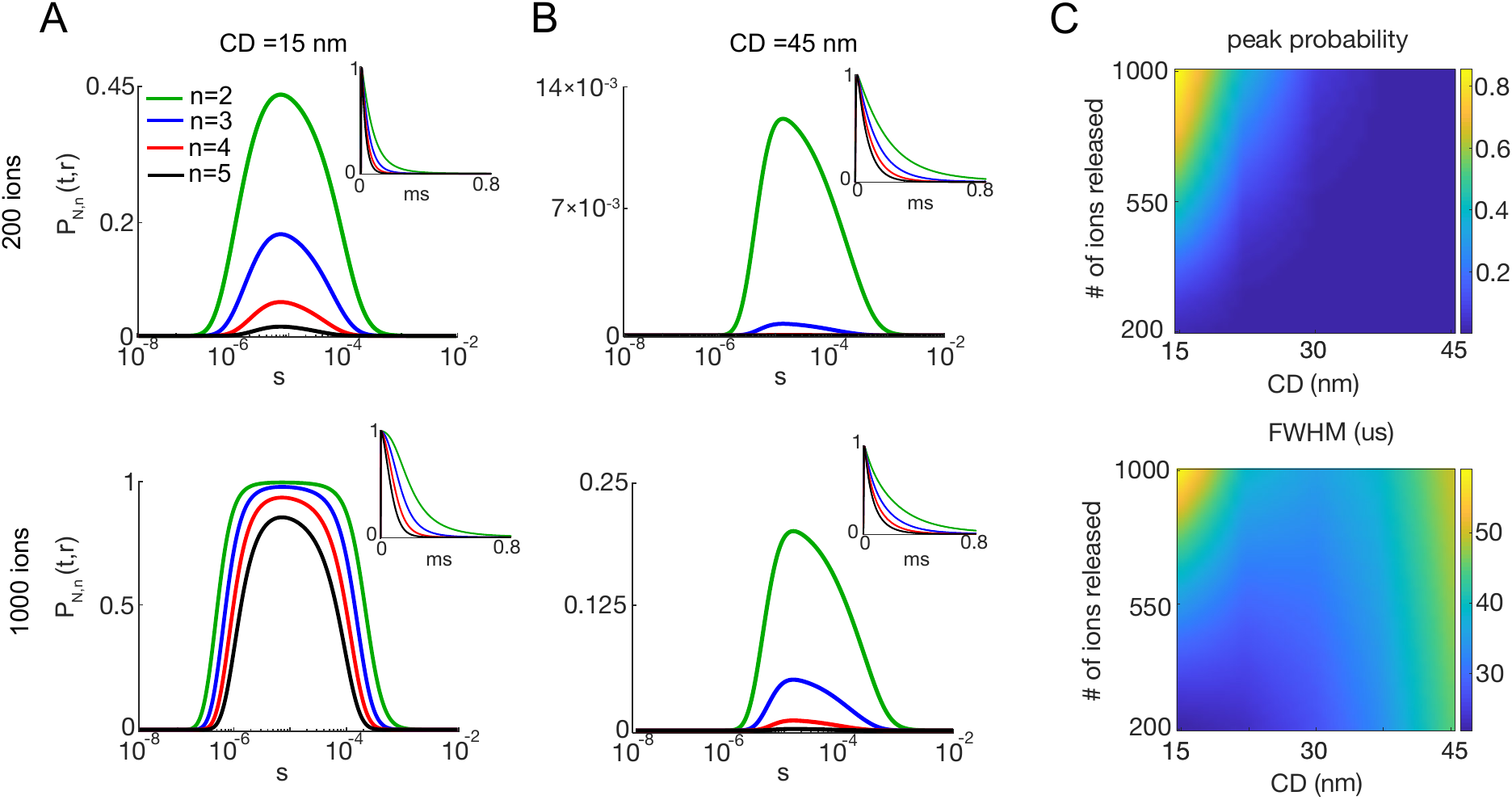}
\caption{\label{fig:fig8}
The occupancy probability $P_{N,n}(t,r)$ of at least $n$ binding sites on the sensor, given various numbers of ions ($N$) and coupling distances (CD). (\textbf{A,B}) $P_{N,n}(t,r)$ for $N = 200$ (1st row) and $N = 1000$ (2nd row) simultaneously released ions for CD of 15 nm (\textbf{A}) and 45 nm (\textbf{B}), for 2 (green), 3 (blue), 4 (red) and 5 (black) binding sites. \textbf{C} Peak (top) and FWHM (bottom)
of the occupancy probability $P_{N,n}(t,r)$ for $n=5$, depending on the number of released calcium ions and coupling distance. }
\end{figure}

Figure \ref{fig:fig8} summarizes the effect of changing the number $n$ of bound calcium ions. As expected, the occupancy probability $P_{N,n}(t,r)$ decreased $n$ for all coupling distances (Fig. \ref{fig:fig8}A,B). The width of this function for larger $n$ is also narrowed. Taking advantage of the analytical approach, we next examine the amplitude and the width of $P_{N,n}(t,r)$ for various number of released calcium ions and coupling distances. One sees that $P_{N,n}(t,r)$ increased with the number $N$ of simultaneously released calcium ions and decreased with increased coupling distance (Fig. \ref{fig:fig8}C). The width of the occupancy probability for $n=5$ shows a decreased sensitivity to increasing $N$ as compared with the single binding site results reported on Fig. \ref{fig:fig6}, where the highest discrepancy between MC and analytical results was observed for large occupancy probabilities, as reflected by the larger difference in the FWHM between MC and analytical curves (Fig. \ref{fig:fig6}A (bottom)). We emphasize that obtaining such a contour plot by Monte Carlo simulations would take approximately 8 months of computation on 250 CPUs. 

In comparison with Figure \ref{fig:fig6}, $P_{N,n}(t,r)$ for $n=5$ stays below the occupancy of 0.5 for up to 500 instantaneously fluxed ions at 15 nm. This is 5-fold more than in the single binding site case. Thus these analytical calculations show that when considering the multi-site binding $Ca^{2+}$ sensor proteins, it is possible to estimate the occupancy of the fully bound sensor without saturation for an influx mediated by a point flux equivalent to that of two simultaneously open channels (500 ions) and a coupling distance of 15 nm. All the simulations taken together, we demonstrate that our first passage-based analytical solution can account for simple multi-site sensors and $Ca^{2+}$ fluxes in the physiological range. However, it cannot account for nonlinearities arising from cooperative alterations in the binding constants during sequential binding events, which could be topic for further study in the future.

\section*{Discussion and Conclusion}

In this work, we introduced an analytical framework for computing the $Ca^{2+}$ occupancy of a target protein sensor for SV fusion following molecular diffusion from a VGCC source. The main novelty of this approach is its ability to account for binding and unbinding kinetics of the sensor in the presence of competing $Ca^{2+}$ buffers. In particular, the unbinding kinetics, which were ignored in former analytical studies, possibly due to theoretical challenges of its implementation \cite{grebenkov2017first}, were shown to be a major factor that shapes the sensor occupancy probability. Our first-passage approach obviates the need to perform computationally intensive MC simulations, and preserves the accuracy in predicting biological stochasticity. 

We first analyzed the sensor occupancy probability $P(t,r)$ for a single $Ca^{2+}$ and its dependence on various diffusion-reaction parameters. In particular, we observed that the peak of $P(t,r)$ was usually at few tens of microseconds, a timescale comparable to experimental measurements of SV fusion times, but a thousand times smaller than the mean first-passage time, which gives a measure of the average rate occurrence of a stochastic event (in our setting, the sensor binding). This is a striking biological example of a process in which the mean first-passage time is misleading, whereas the first-passage time distribution cannot be reduced to its mean. For calculating sensor occupancy in response to a flux of $N$ independent calcium ions, we showed, by comparing to simulations, that $P_N(t,r)$ was accurate provided the peak occupancy probability was less than 0.5. In particular high $P_N(t,r)$ are achieved with large molecular fluxes and short CDs (<20 nm). The peak $P_N(t,r)$ was reduced by including competing $Ca^{2+}$ buffers, thereby increasing the number of $Ca^{2+}$ that could be simulated accurated. While future mathematical studies are needed to account for molecular occlusion that occurs when $P(t,r)$ is high (>0.5), the current analytical solution is accurate for various conditions such as low $Ca^{2+}$ fluxes (corresponding to 0.3 ms single channel currents of 0.3 pA), long CDs (>40 nm) and in the presence of the EFB. Also, the solution was shown to be accurate for the case of $Ca^{2+}$ fluxes via single stochastic VGCC, even at 15 nm. This seems likely due to the lower instantaneous flux that occurs for time-distributed influx during an action potential.

Because the $Ca^{2+}$ dependence of SV fusion is known to be nonlinear (Hill coefficient = 4), we then considered the occupancy of several independent binding sites of the sensor for SV fusion \cite{dittrich2013excess}, which we implemented using a simple combinatorial calculation. This allowed us to calculate the occupancy probability $P_{N,n}(t,r)$ by at least $n$ calcium ions, i.e., the probability that $n$ (or more) calcium ions are simultaneously bound to the sensor at time $t$. 
We showed that the duration of $P_{N,n}(t,r)$ became briefer, meaning that there is a relatively narrow time window during which the SV fusion is possible, as has been observed experimentally \cite{nakamura2015nanoscale}.

Several extensions of the current model and solution are possible. First, one can include multiple SV sensors located in different regions of the synaptic bouton. For this purpose, one can partition the synaptic membrane into ``zones of influence'' around each sensor, as proposed earlier \cite{guerrier2016hybrid,grebenkov2019reversible}; as a calcium ion released within the zone of influence of a given sensor would have more chances to bind to that particular sensor, the binding dynamics within each zone of influence can be studied separately from the others, at least in a first approximation. Second, our analytical solution describes the occupancy probability for a single sensor that can bind a very large (strictly speaking, unlimited) number of $Ca^{2+}$. Accounting for the saturation of the binding sites of the sensor is another important perspective which can be explored by adapting recently proposed models for the dynamics of impatient particles \cite{grebenkov2017first,lawley2019first}. In particular, Lawley and Madrid suggested to model the distribution of the first-passage time onto the target by a mono-exponential function, in which case the number of bound particles can be described by a Markov birth-death process, for which the first-passage time statistics are well known \cite{lawley2019first}. However, the accuracy of such an approximation remains questionable in our setting, 
especially for re-binding steps when the particle unbinds from the sensor and 
diffuses in the bulk until the next binding. Once the bound $Ca^{2+}$ occlusion problem is correctly implemented, it would then be feasible to investigate the cooperative behavior generated by multiple binding sites \cite{naraghi1997linearized}.

In conclusion, our analytical solution allows one to rapidly explore the vast parameters space of the vesicle release process that includes: the binding/unbinding rates on the sensor and on multiple buffers, diffusion coefficients of calcium ions in free and bound states, the sizes of the sensor and the synaptic bouton, the spatial arrangement of multiple VGCCs (or the coupling distance in the case of a single VGCC), the number and the temporal release profile of calcium ions of each VGCC, and the number of binding sites on the sensor. For any physiologically relevant configuration of these parameters, the occupancy probability over a very broad range of timescales (from nanosecond to second) can now be calculated almost instantly, as compared to conventional deterministic or particle-based simulations. The analytical nature of the solution does not generate integration errors from small time steps, which may be critical for conventional numerical techniques at long times. Moreover, the explicit calculation of the occupancy probability for a single calcium ion facilitates incorporation and control of biological stochasticity, particularly for rare events.

More generally, our mean-field approach circumvents the limitations of tracking the diffusion of many thousand of molecules that represent millimolar concentrations found in biology. In fact, the time efficiency of our method is vastly superior to the existing MC alternatives. For instance, the brute force MC simulations with 4~mM of fixed buffer took many hours to complete and other simulations with higher buffer concentrations were not feasible on our computational facilities. In contrast, the analytical solution gave essentially the same results as MC in a fraction of minute, without limitations on molecular concentrations. This opens doors for exploration of complex diffusion-reaction systems that were previously out of reach. 
In this way, one can discover some specific regions of biological interest in the parameters space. For instance,  
Fig. \ref{fig:fig8}C (top) showed the dependence of the peak of the occupancy probability on the number of released calcium ions and on the coupling distance. In this figure, we revealed the region of values of $N$ and CD that produce the occupancy probability that is high enough for the vesicle release. 
Plotting a similar figure with conventional simulations would be extremely long, especially due to such combinations of parameters that produce very small occupancy probabilities. 

At the same time, the derived analytical solution relied on a simplified model the $Ca^{2+}$-sensor binding required for the SV fusion that neglected some biophysical aspects of this sophisticated process (see Methods). Quantifying the respective roles of these ignored aspects and their incorporation into more elaborate models present an important perspective of the present work. Even if such advanced models cannot be solved analytically, yet, they can be analyzed by modern simulators. As a consequence, an improved performance can be achieved by combining the present analytical approach with conventional numerical simulations.

\section*{Methods}

\subsection*{Theoretical model}

(i) {\bf Geometric settings}.  A $Ca^{2+}$ ion (or multiple ions) was injected on the membrane on the distance $r$ from the center of a synaptic terminal (bouton) with radius R, and allowed to diffuse throughout. At the center, we placed a single $Ca^{2+}$
sensor of hemispherical shape and radius $\rho$, a value analogous to an interaction radius (Fig. \ref{fig:chem_syn}A).  
The outer boundary of the bouton (a hemi-spere of radius $R$) is modeled as reflecting, i.e., the flux of $Ca^{2+}$ ions at this boundary is zero.
Note that more elaborate partially reflecting boundary could also be considered to account for $Ca^{2+}$
ions escaping far from the synaptic bouton membrane but we strick
to the reflecting condition here.  
In summary, we consider the
active zone of the shape (Fig. \ref{fig:chem_syn}B)
\begin{equation}  
\Omega_0= \{ \x=(x,y,z) \in \mathbb{R}^3 :\rho<|\x|<R,~ z>0 \} .
\end{equation}
Importantly, we neglect the presence of the synaptic vesicle whose
reflecting boundary might hinder the motion of $Ca^{2+}$ ions; in fact,
it has been shown by Monte Carlo simulations that in physiological
conditions, the synaptic vesicle does not influence the single vesicle
release probability \cite{nakamura2018variations}.

(ii) {\bf $Ca^{2+}$ ions} are modeled as independent point-like
diffusing particles that undergo Brownian motion with diffusion
coefficient $D_0$ in the region $\Omega_0$ between the boundaries of
sensor and active zone; in particular, the charge of $Ca^{2+}$ ions is
ignored due to bulk screening of electrostatic interactions.  A
$Ca^{2+}$ ion source (VGCC) was set at a fixed distance $r-\rho$ from
the sensor (we discuss below how to deal with multiple sources).

(iii) {\bf Buffers} are modeled as co-existing continuous homogeneous
reactive media that can bind, transport, and release $Ca^{2+}$ ions;
their functioning is assumed to be in a linear regime (i.e., low
occupancy),
i.e. the exchange between the free $Ca^{2+}$ state (denoted
by index $0$) and the bound state with the $i$-th buffer (denoted by
index $i$) occurs through the standard first-order kinetics, with the
exchange rates $k_{0i}$ and $k_{i0}$; the $Ca^{2+}$ ion in a bound state
diffuses with the diffusion coefficient $D_i$ but cannot bind to the
sensor.  Under the assumption of a homogeneous reactive medium, the
``binding rate'' can be expressed as $k_{0i} = k_{{\rm on},i} \,
c_i$, where $k_{{\rm on},i}$ is the conventional binding constant and
$c_i$ is the concentration of the $i$-th buffer. 

(iv) {\bf Sensor kinetics}. In a basic setting, we consider a sensor with a single
binding site, its kinetics is determined by $\kon$ and $\koff$ binding
rate constants. In most cases (unless stated otherwise), we used
reaction rate constants identical to the reaction rate constant of the
first binding site from the 5 state sensor model
\cite{wang2008synaptic}  (see also below). When the $Ca^{2+}$ ion reaches the surface of
the sensor it can be reflected from it or bind to it, the random
choice of either depending on the sensor binding constant $\kon$
\cite{grebenkov2003,grebenkov2019c}. When bound, the $Ca^{2+}$ ion
remains in this state for a random time $\tau$ distributed by an
exponential law
\begin{equation}  \label{eq:Phi}
\Phi(t) = \P\{\tau>t\} = \exp(-\koff \, t)  , 
\end{equation}
$1/\koff$ being the mean waiting time before unbinding reaction.
After the unbinding from the sensor, the $Ca^{2+}$ ion resumes its
diffusion in the intrasynaptic region $\Omega_0$ until the next
binding event.

As any model, our theoretical description is based on assumptions, in which some reactions were simplified and others neglected.
For instance, we ignored $Ca^{2+}$ extrusion mechanisms that exist 
in the synapse to return the $Ca^{2+}$ level to baseline within 
hundreds of milliseconds \cite{helmchen1997calcium}. Even though these
mechanisms could affect our results for timescales greater than $t \gtrsim 0.1$~seconds,
they are irrelevant in the microsecond/millisecond range, at which
the occupancy probability is maximal and thus the SV fusion would
most likely occur. We note that these mechanisms could be 
introduced into our model via reversible reactions with an additional 
(artificial) buffer. We also neglected the baseline $Ca^{2+}$ level 
that plays important roles, e.g. for spontaneous SV release
\cite{lou2005allosteric}. This residual level of resting $Ca^{2+}$ ions can be
included into our diffusion-reaction equations as an appropriate  
initial condition (e.g., with a uniform concentration) or via a 
source term on the outer boundary of the synaptic bouton. The 
effect of this baseline level onto the occupancy probability and
the consequent spontaneous SV release can be further investigated. 
Perhaps, the most significant simplification is the assumption 
of unlimited binding capacity of the sensor and the consequent 
consideration of multiple $Ca^{2+}$ binding events as independent
(see below). This simplification allows for the detailed examination of spatio-temporal relationship between nanoscale $[Ca^{2+}]$ gradients and their ability to drive nonlinear SV fusion reactions – both critical features of synaptic transmission.

\subsection*{Sensor occupancy probability}

We are interested in computing the so-called occupancy probability
$P(t,\x)$ that a particle (here, a $Ca^{2+}$ ion), started from a fixed
point $\x$ at time $0$, is at the bound state on the sensor at a later
time $t$.  Due to the sensor kinetics, the particle can undergo
numerous binding/unbinding events up to time $t$.  To account for
these events, we introduce an auxiliary probability density
$\psi_n(t,\x)$ of the $n$-th binding at time $t$.  These densities can
be obtained via recurrent functional relations.  In fact, the
independence between the time spent in the bound state on the sensor,
and the time of a bulk excursion after unbinding, implies
\begin{equation}
\label{eq:psin_rec}
\psi_n(t,\x) = \int\limits_0^t dt_1 \int\limits_{t_1}^t dt_2 \, \psi_{n-1}(t_1,\x) \, \phi(t_2-t_1) \, \psi(t-t_2) ,
\end{equation}
where $\phi(t) = \koff\, \exp(-\koff \, t)$ is the probability density
of the exponential waiting time in the sensor-bound state, and $\psi(t)$ is
the probability density of re-binding at time $t$ after the release at
time $0$.  This is a standard renewal relation, which states that,
after the $(n-1)$-th binding of the particle at some time $t_1$ (with
the density $\psi_{n-1}(t_1,\x)$), the particle remains bound during
time $t_2-t_1$ and unbinds at time $t_2$ (with the density
$\phi(t_2-t_1)$), diffuses in the bulk during time $t-t_2$ and
re-binds at time $t$ (with the density $\psi(t-t_2)$).  Since the
intermediate binding/unbinding events may occur at any times between
$0$ and $t$, one has to integrate over $t_1$ and $t_2$.  The integral
relation (\ref{eq:psin_rec}) is reduced to a product in the Laplace
space, i.e.,
\begin{equation}
\tilde{\psi}_n(p,\x) = \tilde{\psi}_{n-1}(p,\x) \, \tilde{\phi}(p) \, \tilde{\psi}(p)  
 = \tilde{\psi}_1(p,\x) \bigl[\tilde{\phi}(p) \, \tilde{\psi}(p)\bigr]^{n-1},   
\end{equation}
where tilde denotes Laplace-transformed quantities, e.g.,
\begin{equation}
\tilde{\psi}_n(p,\x) = \int\limits_0^\infty dt \, e^{-pt} \, \psi_n(t,\x).
\end{equation}

The probability of a particle to be in the bound state at time $t$ can
be expressed as follows
\begin{equation}
P(t,\x) = \sum\limits_{n=1}^\infty \int\limits_0^t dt' \, \psi_n(t',\x) \, \Phi(t-t') .
\end{equation}
In this infinite sum, the $n$-th term is the probability that after
the $n$-th binding at time $t'$ (with the density $\psi_n(t',\x)$),
the particle remains at the bound state for time $t-t'$ (with the
probability $\Phi(t-t')$ given by Eq. (\ref{eq:Phi})).  This relation
simply reflects the fact that the particle, which is at the bound
state at time $t$, has experienced either $1$, or $2$, $\ldots$ or
$n$, or $\ldots$ binding events.  In the Laplace space, we get
\begin{equation} 
\label{eq:tildeP}
\tilde{P}(p,\x) = \sum\limits_{n=1}^\infty \tilde{\psi}_n(p,\x)\, \tilde{\Phi}(p) 
 = \tilde{\psi}_1(p,\x) \biggl(1 - \tilde{\phi}(p) \, \tilde{\psi}(p)\biggr)^{-1} \tilde{\Phi}(p) . 
\end{equation}
The three factors in the product have a clear interpretation: the
first arrival and binding onto the sensor, multiple re-binding events
on the sensor, and waiting after the last re-binding.  For exponential
waiting times, one has $\tilde{\Phi}(p) = 1/(p+\koff)$ and
$\tilde{\phi}(p) = \koff/(p+\koff)$, and thus we finally get

\begin{equation} 
\label{eq:tildePP}
\tilde{P}(p,\x) = \tilde{\psi}_1(p,\x) \, \biggl(p + \koff (1-\tilde{\psi}(p))\biggr)^{-1} .
\end{equation}

The inverse Laplace transform of Eq. (\ref{eq:tildePP}) allows one to
return to the time domain to get $P(t,\x)$.  In this way, the
probability $P(t,\x)$ is reduced to the analysis of the ``elementary''
diffusion step -- the binding to the sensor -- that determines both
$\tilde{\psi}_1(p,\x)$ and $\tilde{\psi}(p)$.  We emphasize that
Eq. (\ref{eq:tildePP}), written in terms of survival probabilities, is
well known for describing reversible kinetics in chemical physics, see
\cite{agmon1990,prustel2013} and references therein.

\subsection*{Distribution of the first-binding time}

We start by noting that the symmetry of the considered domain
$\Omega_0$ allows one to effectively remove the synaptic bouton
membrane at $z = 0$ and thus replace $\Omega_0$ by a simpler spherical
layer

\begin{equation}
\Omega = \{ \x = (x,y,z) \in {\mathbb R}^3 ~:~ \rho < |\x| < R \}.
\end{equation}
In other words, the distribution of first-passage times computed in
$\Omega_0$ is identical to that computed in $\Omega$.  The advantage
of the latter domain is that it is rotation invariant so that the
problem can be reduced to one-dimensional radial part, as discussed
below.  When there is no buffer, the computation of the first-passage
time to a target is rather standard
\cite{redner2001,crank1975,carslaw1959} but technically involved in
the case of a spherical layer \cite{grebenkov2018}.  Accounting for
buffers presents one of the major challenges and originalities of this
work.  Note that our approach generalizes some earlier results for
two-channel diffusion \cite{godec2017}.

We investigate the model with $M$ distinct buffers by using an
$(M+1)$-state switching diffusion model: the $Ca^{2+}$ ion can be either
in a free state ($0$) or in a buffer-bound state ($i$), with
$i=1,\ldots,M$.  Given that the buffers are modeled as continuous and
homogeneous media, a transition from the state $i$ to the state $j$
happens spontaneously, with a given rate $k_{ij}$ (see Section I of
the SI for a formal definition of the model).
A general scheme for studying first-passage times for switching
diffusions was recently developed in \cite{grebenkov2019}.  We
introduce $(M+1)$ survival probabilities $S_i(t,\x)$ for a $Ca^{2+}$ ion
started at $\x$ in the state $i$ to be unbound from the sensor until
time $t$.  These probabilities satisfy $(M+1)$ coupled backward
Fokker-Planck (or Kolmogorov) equations \cite{yin2010}:
\begin{equation}
\label{eq:Si_diffusion}
\partial_t S_i = D_i \Delta S_i + \sum\limits_{j=0}^M k_{ij} (S_j - S_i)   \qquad (i = 0,1,\ldots,M),
\end{equation}
subject to the initial condition: $S_i(0,\x) = 1$.  Here $D_i$ is the
diffusion coefficient of the $Ca^{2+}$ ion in the state $i$, $\Delta$ is
the Laplace operator, and we set $k_{ii} = 0$ to simplify notations.
We recall that there is no direct $Ca^{2+}$ ion exchange between bound
states:
\begin{equation}  \label{eq:kij_cond}
k_{ij} = 0 \qquad (1\leq i,j\leq M).
\end{equation}
In other words, any exchange between the states $i$ and $j$ occurs
through the free state $0$.  The last term in
Eqs. (\ref{eq:Si_diffusion}) describes transitions between states $i$
and $j$. 

Equations (\ref{eq:Si_diffusion}) should be completed by boundary
conditions at the inner sphere at $|\x| = \rho$ (the sensor) and the
outer sphere at $|\x| = R$ (the frontier of the active zone).  The
outer reflecting boundary simply confines the $Ca^{2+}$ ions within the
active zone, i.e., it ensures that there is no flux of $Ca^{2+}$ ions
across this boundary:
\begin{equation}
\label{eq:Neumann_outer}
- D_i \partial_n S_i(t,\x) =  0 \quad (|\x|=R, ~ i=0,1,\ldots,M),
\end{equation}
where $\partial_n$ is the normal derivative directed outwards the
domain.  Since the $Ca^{2+}$ ions in bound states cannot bind to the
sensor, the same Neumann boundary condition is imposed at the inner
sphere:
\begin{equation}
- D_i \partial_n S_i(t,\x) =  0 \quad (|\x|=\rho, ~ i=1,\ldots,M).
\end{equation}
Finally, the calcuim ions in the free state can bind to the sensor
that implies the Robin boundary condition
\begin{equation}
- D_0 \partial_n S_0(t,\x) = \frac{\kon S_0(t,\x)}{N_A (4\pi \rho^2)}  \quad (|\x|=\rho) .
\end{equation}
It is obtained by equating the net diffusive flux at the sensor
(left-hand side) to the reactive flux (right-hand side) controlled by
the reaction constant $\kon$, where $N_A$ is the Avogadro number, and
$4\pi \rho^2$ is the surface area of the sensor
\cite{shoup1982,lauffenburger1993}.  We emphasize that the presence of
buffers has two effects: change in the diffusion coefficient and
impossibility of a buffer-bound $Ca^{2+}$ ion to bind to the sensor.
Since the calcuim ions are released in the free state, we are
interested exclusively in $S_0(t,\x)$.  However, finding this
probability requires solving the coupled system of equations for all
$S_i$.  Note that $1 - S_0(t,\x)$ describes the fraction of $Ca^{2+}$
ions that have been bound to the sensor up to time $t$.  This is the
cumulative probability distribution for the first-binding time.  In
particular, its probability density reads
\begin{equation}
\label{eq:psi1_S0}
\psi_1(t,\x) = \partial_t \bigl(1 - S_0(t,\x)\bigr) = - \partial_t S_0(t,\x) .
\end{equation}
Due to the rotational invariance of the problem, the probabilities
$S_i(t,\x)$ and the probability density $\psi_1(t,\x)$, written in
spherical coordinates, depend only on the radial coordinate $r =
|\x|$.  From now on, we replace $\x$ by $r$.

The solution of the system (\ref{eq:Si_diffusion}) of coupled partial
differential equations is detailed in Section II of the SI.  In a
nutshell, the Laplace transform reduces these equations to a system of
ordinary differential equations with respect to the radial coordinate
$r$ that is then solved by standard methods.  Once the solution is
found, one gets from Eq. (\ref{eq:psi1_S0})
\begin{equation}
\label{eq:tilde_psi1}
\tilde{\psi}_1(p,r) = 1 - p \tilde{S}_0(p,r) .
\end{equation}
In addition, as a bulk excursion after the unbinding event starts at
the sensor surface, $r = \rho$, one has
\begin{equation}
\tilde{\psi}(p) = \tilde{\psi}_1(p, \rho) .
\end{equation}
As a consequence, the knowledge of $\tilde{S}_0(p,r)$ yields both
dynamical characteristics, $\tilde{\psi}_1(p,r)$ and
$\tilde{\psi}(p)$, that determine the Laplace-transformed probability
$\tilde{P}(p,r)$ according to Eq. (\ref{eq:tildePP}).

The last step for getting $P(t,r)$ in time domain requires the inverse
Laplace transform of $\tilde{P}(p,r)$.  This is performed by
determining the poles of this function and applying the residue
theorem.  When the poles are simple, the occupancy probability admits
the following exact representation:
\begin{equation}  \label{eq:Pt}
P(t,r) = P_\infty + \sum\limits_{n=1}^\infty \exp\bigl(- \alpha_n^2 D_0 t/\rho^2 \bigr) \sum\limits_{j=0}^M b_n^{(j)} \, u(\alpha_n^{(j)},r)  ,
\end{equation}
where
\begin{equation}
\label{eq:ur}
u(\alpha, r) = \frac{\rho \, \sin\bigl(\alpha \frac{R-r}{\rho}\bigr) - R\alpha \cos\bigl(\alpha \frac{R-r}{\rho}\bigr)}{r} \,,
\end{equation}
and 
\begin{equation}  \label{eq:Pinf}
P_\infty = \biggl(1 + \koff \frac{4\pi (R^3 - \rho^3)N_A}{3\kon} \biggl(1 + \sum\limits_{j=1}^M \frac{k_{0j}}{k_{j0}}\biggr) \biggr)^{-1} 
\end{equation}
is the steady-state limit.  The coefficients $b_n^{(j)}$ and
$\alpha_n^{(j)}$ are determined by exact but complicated formulas
provided in the SI, whereas $\alpha_n$ are found as strictly positive
solutions of some trigonometric equation (provided in the SI).  For
instance, in the simplest case of no buffer ($M = 0$), we obtained in
Section III of the SI:
\begin{equation}
\label{eq:bn_M0}
\alpha_n^{(0)} = \alpha_n, \qquad 
b_n^{(0)} = \frac{2 \mu}{\sin(\alpha_n \beta) \bigl(\alpha_n^2 w_1 + w_2 \bigr) 
 + \alpha_n \cos(\alpha_n\beta) \bigl(\alpha_n^2 w_3 + w_4 \bigr)} ,
\end{equation}
where
\begin{subequations}
\begin{align}
w_1 & = 4(1+\beta) + \beta(\beta+\mu(1+\beta)) , \\
w_2 & = 2(1+\mu-\lambda(1+\beta)) - \lambda \beta^2 , \\
w_3 & = \beta(1+\beta) ,\\
w_4 & = \beta(1 + \mu - \lambda(1+\beta)) - 3(\beta + \mu(1+\beta))  ,
\end{align}
\end{subequations}
with dimensionless parameters
\begin{equation}
\beta = (R - \rho)/\rho, \qquad  \lambda = \koff \rho^2/D_0 , \qquad  \mu = \kon/(4\pi \rho D_0 N_A), 
\end{equation}
and $\alpha_n$ are strictly positive solutions of the trigonometric
equation
\begin{equation}
\label{eq:alpha_M0}
\sin(\alpha_n \beta) = \frac{\bigl[\alpha_n^2(\beta+\mu(1+\beta)) - \lambda \beta\bigr] \alpha_n \cos(\alpha_n \beta)}
{\alpha_n^4(1+\beta) + \alpha_n^2(1+\mu - \lambda(1+\beta)) - \lambda}   \qquad (n=1,2,\ldots).
\end{equation}
For a single buffer ($M = 1$), we also derived explicit formulas but
they are much more cumbersome (see Section IV of the SI).  Even though
analytical calculations become prohibitively complicated for $M > 1$,
numerical computations based on our analytical solution remain fast
and accurate.

The exact solution (\ref{eq:Pt}) is the main analytical result of the
paper.  Although this solution may look cumbersome and involves some
numerical steps (truncation of the infinite series, numerical
computation of the coefficients, etc), its explicit form allows for
both analytical and numerical investigation of the occupancy
probability $P(t,r)$.

\subsection*{Instant calcium influx}

If $N$ independent ions are released simultaneously from the same
fixed position $r$, then the single binding site
occupancy probability can be computed as a probability of at least one out of $N$
ions being bound to the sensor, according to the formula:
\begin{equation}  \label{eq:PN}
   P_N(t,r)= 1-(1-P(t,r))^N,
\end{equation}
where $P(t,r)$ is the occupancy probability of single binding site by
a single ion. The above formula relies on the assumption that the sensor has an unlimited binding capacity. In other words, the sensor can simultaneously bind $1,2,3,\ldots,N$ calcium ions, and its binding rate $\kon$ does not depend on the number of already bound ions. This assumption was crucial to be able to consider the reaction kinetics of $N$ calcium ions independently from each other and thus to reduce the original very complicated "$N$-body problem" to a simple relation (\ref{eq:PN}) involving only the occupancy probability $P(t,r)$ for a single ion. In statistical physics, this would correspond to a mean-field approximation.

Similarly, one can compute the occupancy probability $P_{N,n}(t,r)$ for at least $n$ calcium ions bound simultaneously to the sensor at time $t$ as
\begin{equation}  \label{eq:PNn}
P_{N,n}(t,r) = 1 - \sum_{k=0}^{n-1} \binom{N}{k} \, [P(t,r)]^k \, (1-P(t,r))^{N-k} \,,
\end{equation}
where $\binom{N}{k}$ is the binomial coefficient. For $n = 1$, this formula is reduced to Eq. (\ref{eq:PN}) for $P_N(t,r)$. If there was no unbinding kinetics from the sensor (i.e., if $\koff = 0$), $1-P_{N,n}(t,r)$ could be interpreted as the "survival" probability that there were no $n$ bound calcium ions up to time $t$. This survival probability would then determine the probability density of the first moment ${\mathbb T}_n$ when $n$ calcium ions are simultaneously bound to the sensor. As calcium ions could not unbind, this is precisely the first moment $\hat{{\mathbb T}}_n$ of binding of the $n$-th calcium ion: $\hat{{\mathbb T}}_n = {\mathbb T}_n$. However, the possibility of unbinding (i.e., $\koff > 0$) transforms this equality into inequality, $\hat{{\mathbb T}}_n < {\mathbb T}_n$, and makes the computation of 
${\mathbb T}_n$ a challenging open problem. This mathematical difficulty was probably one of the reasons why the unbinding kinetics was ignored in former theoretical works on this topic. Here, we made an important step toward a more realistic analytical model by incorporating the unbinding kinetics into the occupancy probability $P_{N,n}(t,r)$ which can be seen as a proxy for the likelihood of the SV fusion at time $t$.

In the MC simulations $P_N(t,r)$ was computed as the probability of
finding a single ion bound to the sensor, given that $N$ ions were
released at the same time from the same position.

\subsection*{Calcium influx through single VGCC}

Single VGCC was modeled as a three-step Hodgkin-Huxley process
\cite{hodgkin1952quantitative} (see also Section V the SI), the model and parameters are taken Ref. \cite{rebola2019distinct}.  This model reproduces the single channel characteristics that were measured previously: opening
probability 0.3, maximum of the single channel current 0.3 pA, full
width half maximum of the single channel current 250 $\mu$s.

Simulation of $Ca^{2+}$ influx through the channel was done using MC
tool, with 1000 trials. For the $j$-th trial, we stored the random times $t_{1}^{(j)}, \ldots, t_{N}^{(j)}$ when $N$ ions ``entered'' the system. Given the instances of ion appearance we
calculated the probability of sensor occupancy for this trial
using Poisson binomial distribution at each point:
\begin{equation} 
   P_{trial}^{(j)}(t,r) = 1 - \prod\limits_{i=1}^N (1-P(t+t_i^{(j)},r)).
\end{equation}
Then these probabilities were averaged among the trials:
\begin{equation} 
   P_{AP}(t,r) = \frac{1}{1000} \sum_{j=1}^{1000}  P_{trial}^{(j)}(t,r) .
   \label{eq3}
\end{equation}
More generally, if the ions entered from different VGCC channels, one could use $P(t+t_i^{(j)},r_i^{(j)})$ with the appropriate location $r_i^{(j)}$ of the source of the $i$-th ion in the $j$-th trial. In this way, one can easily implement sophisticated spatio-temporal characteristics of the $Ca^{2+}$ ions release.

The analytical solution from Eq. (\ref{eq3}) can be compared to the direct estimate of this probability from Monte Carlo simulations. For each simulation trial, we computed the moments of binding and unbinding of calcium ions that determine the number of bound calcium ions at time $t$, $N^{(j)}(t,r)$, at the trial $j$. The average of these functions over all trials is the direct estimate of the probability:
\begin{equation}
    P_{AP}^{MC}(t,r) = \frac{1}{1000} \sum_{j=1}^{1000}  N^{(j)}(t,r) .
\end{equation}
The comparison between $P_{AP}(t,r)$ and $P_{AP}^{MC}(t,r)$ is shown on Fig. \ref{fig:fig7}B.

\subsection*{Stochastic simulations}

For verification of analytical results we use particle-based stochastic numerical simulations (MCell software \cite{kerr2008fast}). In MCell diffusion of individual molecules is modeled using Brownian dynamics, while chemical reactions occur due to the collision of molecules and follow Poisson distribution. All the parameters for simulations are identical to the parameters of analytical solution. The presynaptic domain of radius 300 nm and sensors are modeled as spheres, intersected by a reflecting plane in the origin (Fig. \ref{fig:chem_syn}B). The sensor is located in the origin of the volume and has a radius of 5 nm. 
Depending on the context of the simulation, the $Ca^{2+}$ input, number of calcium channels and the distance between the sensor and calcium channels were manipulated; for instance, for the computation of the occupancy probability by a single ion, it was released at time 0 from a single source. Each time the particle hits the sensor was recorded. The first-passage time distribution was computed based on the recorded times. 

To compute occupancy probabilities we stored the time instances of the reaction between $Ca^{2+}$ ion and the sensor, then the number of binding events at each time instance was divided by the total number of trials.   
The interaction range between two particles was set to 5 nm, the time step was chosen to be 5 ns.

\subsection*{Deterministic simulations}

The release rates can be simulated using a 5-state model of $Ca^{2+}$ triggered vesicle fusion \cite{wang2008synaptic}:
\begin{equation}\label{eq:5state}
V_0 \mathrel{\mathop{\rightleftarrows}^{5 \kon}_{\koff b^0}} V_1 \mathrel{\mathop{\rightleftarrows}^{4 \kon}_{2 \koff b^1}} 
V_2 \mathrel{\mathop{\rightleftarrows}^{3 \kon}_{3 \koff b^2}} V_3 
\mathrel{\mathop{\rightleftarrows}^{2 \kon}_{4 \koff b^3}} V_4 
\mathrel{\mathop{\rightleftarrows}^{\kon}_{5 \koff b^4}} V_5 
\mathrel{\mathop{\rightarrow}^{\gamma}} F ,
\end{equation}
where $V_i$ denote the binding states of the sensor (i.e., the sensor
with $i$ calcium ions bound, and $V_0$ meaning the unbound state), and
$F$ is the fused state of the vesicle. The conventional values of the 
parameters are \cite{wang2008synaptic}: $\kon = 127~\konU$, 
$\koff=15.7~\koffU$, $b=0.25$, $\gamma=6~\koffU$. We used these values 
for plotting Fig. \ref{fig:fig1}. For this purpose, the system of ordinary differential equations describing this model was integrated using forward Euler scheme in a custom Matlab routine. The input $Ca^{2+}$ transients are results of the spatial deterministic simulations for the channel-vesicle arrangement as in Fig. \ref{fig:fig1}, provided by Yukihiro Nakamura. The sensor binding sites were assumed not to alter the free calcium due to their small number.

For MC simulations of single binding site occupancies, which were used to compare with analytical solutions, we used five times $\kon$ ($127~\konU$; see Table \ref{tab:param}).


\begin{thebibliography}{52}
\urlstyle{rm}
\expandafter\ifx\csname url\endcsname\relax
  \def\url#1{\texttt{#1}}\fi
\expandafter\ifx\csname urlprefix\endcsname\relax\def\urlprefix{URL }\fi
\expandafter\ifx\csname doiprefix\endcsname\relax\def\doiprefix{DOI: }\fi
\providecommand{\bibinfo}[2]{#2}
\providecommand{\eprint}[2][]{\url{#2}}

\bibitem{alberts2008specialized}
\bibinfo{author}{Alberts, B.} \emph{et~al.}
\newblock \bibinfo{title}{Specialized tissues, stem cells,
  and tissue renewal}.
\newblock {\emph{Molecular biology of the cell}. 5th ed. New York:
  Garland Science, Taylor \& Francis Group}  (\bibinfo{year}{2008}).

\bibitem{berridge2003calcium}
\bibinfo{author}{Berridge, M.~J.}, \bibinfo{author}{Bootman, M.~D.} \& \bibinfo{author}{Roderick, H.~L.}
\newblock \bibinfo{title}{Calcium: calcium signalling:
  dynamics, homeostasis and remodelling}.
\newblock \bibinfo{journal}{\emph{Nature Reviews Molecular Cell Biology}}
  \textbf{\bibinfo{volume}{4}}, \bibinfo{pages}{517} (\bibinfo{year}{2003}).

\bibitem{eggermann2012nanodomain}
\bibinfo{author}{Eggermann, E.}, \bibinfo{author}{Bucurenciu, I.},
  \bibinfo{author}{Goswami, S.~P.} \& \bibinfo{author}{Jonas, P.}
\newblock \bibinfo{title}{Nanodomain coupling between Ca2+
  channels and sensors of exocytosis at fast mammalian synapses}.
\newblock \bibinfo{journal}{\emph{Nature Reviews Neuroscience}}
  \textbf{\bibinfo{volume}{13}}, \bibinfo{pages}{7--21} (\bibinfo{year}{2012}).

\bibitem{nakamura2015nanoscale}
\bibinfo{author}{Nakamura, Y.} \emph{et~al.}
\newblock \bibinfo{title}{Nanoscale distribution of
  presynaptic Ca2+ channels and its impact on vesicular release during
  development}.
\newblock \bibinfo{journal}{\emph{Neuron}} \textbf{\bibinfo{volume}{85}},
  \bibinfo{pages}{145--158} (\bibinfo{year}{2015}).

\bibitem{roberts1994localization}
\bibinfo{author}{Roberts, W.~M.}
\newblock \bibinfo{title}{Localization of calcium signals by
  a mobile calcium buffer in frog saccular hair cells}.
\newblock \bibinfo{journal}{\emph{Journal of Neuroscience}}
  \textbf{\bibinfo{volume}{14}}, \bibinfo{pages}{3246--3262}
  (\bibinfo{year}{1994}).

\bibitem{matveev2004facilitation}
\bibinfo{author}{Matveev, V.}, \bibinfo{author}{Zucker, R.~S.} \&
  \bibinfo{author}{Sherman, A.}
\newblock \bibinfo{title}{Facilitation through buffer
  saturation: constraints on endogenous buffering properties}.
\newblock \bibinfo{journal}{\emph{Biophysical Journal}}
  \textbf{\bibinfo{volume}{86}}, \bibinfo{pages}{2691--2709}
  (\bibinfo{year}{2004}).

\bibitem{dittrich2013excess}
\bibinfo{author}{Dittrich, M.} \emph{et~al.}
\newblock \bibinfo{title}{An excess-calcium-binding-site
  model predicts neurotransmitter release at the neuromuscular junction}.
\newblock \bibinfo{journal}{\emph{Biophysical Journal}}
  \textbf{\bibinfo{volume}{104}}, \bibinfo{pages}{2751--2763}
  (\bibinfo{year}{2013}).

\bibitem{andrews2010detailed}
\bibinfo{author}{Andrews, S.~S.}, \bibinfo{author}{Addy, N.~J.},
  \bibinfo{author}{Brent, R.} \& \bibinfo{author}{Arkin, A.~P.}
\newblock \bibinfo{title}{Detailed simulations of cell
  biology with smoldyn 2.1}.
\newblock \bibinfo{journal}{\emph{PLoS Computational Biology}}
  \textbf{\bibinfo{volume}{6}}, \bibinfo{pages}{e1000705}
  (\bibinfo{year}{2010}).

\bibitem{modchang2010comparison}
\bibinfo{author}{Modchang, C.} \emph{et~al.}
\newblock \bibinfo{title}{A comparison of deterministic and
  stochastic simulations of neuronal vesicle release models}.
\newblock \bibinfo{journal}{\emph{Physical Biology}}
  \textbf{\bibinfo{volume}{7}}, \bibinfo{pages}{026008} (\bibinfo{year}{2010}).

\bibitem{blackwell2006efficient}
\bibinfo{author}{Blackwell, K.~T.}
\newblock \bibinfo{title}{An efficient stochastic diffusion
  algorithm for modeling second messengers in dendrites and spines}.
\newblock \bibinfo{journal}{\emph{Journal of Neuroscience Methods}}
  \textbf{\bibinfo{volume}{157}}, \bibinfo{pages}{142--153}
  (\bibinfo{year}{2006}).

\bibitem{chen2017parallel}
\bibinfo{author}{Chen, W.} \& \bibinfo{author}{De~Schutter, E.}
\newblock \bibinfo{title}{Parallel steps: large scale
  stochastic spatial reaction-diffusion simulation with high performance
  computers}.
\newblock \bibinfo{journal}{\emph{Frontiers in Neuroinformatics}}
  \textbf{\bibinfo{volume}{11}}, \bibinfo{pages}{13} (\bibinfo{year}{2017}).

\bibitem{donovan2016unbiased}
\bibinfo{author}{Donovan, R.~M.}, \bibinfo{author}{Tapia, J.-J.}, \bibinfo{author}{Sullivan, D.~P.}, \bibinfo{author}{Faeder, J. R.}, 
	\bibinfo{author}{Murphy, R. F.}, \bibinfo{author}{Dittrich, M.}, \& \bibinfo{Zuckerman, D.~M.}
\newblock \bibinfo{title}{Unbiased rare event sampling in spatial stochastic systems biology 
  models using a weighted ensemble of trajectories}.
\newblock \bibinfo{journal}{\emph{PLoS Computational Biology}}
  \textbf{\bibinfo{volume}{12}}, \bibinfo{pages}{e1004611} (\bibinfo{year}{2016}).



\bibitem{naraghi1997linearized}
\bibinfo{author}{Naraghi, M.} \& \bibinfo{author}{Neher, E.}
\newblock \bibinfo{title}{Linearized buffered Ca2+ diffusion
  in microdomains and its implications for calculation of [Ca2+] at the mouth
  of a calcium channel}.
\newblock \bibinfo{journal}{\emph{Journal of Neuroscience}}
  \textbf{\bibinfo{volume}{17}}, \bibinfo{pages}{6961--6973}
  (\bibinfo{year}{1997}).

\bibitem{nakamura2018variations}
\bibinfo{author}{Nakamura, Y.}, \bibinfo{author}{Reva, M.} \&
  \bibinfo{author}{DiGregorio, D.~A.}
\newblock \bibinfo{title}{Variations in Ca2+ influx can alter
  ca2+-chelator-based estimates of ca2+ channel-synaptic vesicle coupling
  distance}.
\newblock \bibinfo{journal}{\emph{Journal of Neuroscience}}
  \bibinfo{pages}{2061--17} (\bibinfo{year}{2018}).

\bibitem{holcman2014}
\bibinfo{author}{Holcman, D.} \& \bibinfo{author}{Schuss, Z.}
\newblock \bibinfo{title}{The narrow escape problem}.
\newblock \bibinfo{journal}{\emph{SIAM Reviews}} \textbf{\bibinfo{volume}{56}},
  \bibinfo{pages}{213--257} (\bibinfo{year}{2014}).

\bibitem{metzler2014}
\bibinfo{author}{Metzler, R.}, \bibinfo{author}{Oshanin, G.} \&
  \bibinfo{author}{Redner, S.~E.}
\newblock \emph{\bibinfo{title}{First-Passage Phenomena and Their
  Applications}} (\bibinfo{publisher}{World Scientific Press},
  \bibinfo{year}{2014}).

\bibitem{grebenkov2017}
\bibinfo{author}{Grebenkov, D.~S.}
\newblock \bibinfo{title}{First passage times for multiple
  particles with reversible target-binding kinetics}.
\newblock \bibinfo{journal}{\emph{Journal of Chemical Physics}}
  \textbf{\bibinfo{volume}{147}}, \bibinfo{pages}{134112}
  (\bibinfo{year}{2017}).

\bibitem{guerrier2016hybrid}
\bibinfo{author}{Guerrier, C.} \& \bibinfo{author}{Holcman, D.}
\newblock \bibinfo{title}{Hybrid markov-mass action law model
  for cell activation by rare binding events: Application to calcium induced
  vesicular release at neuronal synapses}.
\newblock \bibinfo{journal}{\emph{Scientific Reports}}
  \textbf{\bibinfo{volume}{6}} (\bibinfo{year}{2016}).

\bibitem{grebenkov2017first}
\bibinfo{author}{Grebenkov, D.~S.}
\newblock \bibinfo{title}{First passage times for multiple
  particles with reversible target-binding kinetics}.
\newblock \bibinfo{journal}{\emph{Journal of Chemical Physics}}
  \textbf{\bibinfo{volume}{147}}, \bibinfo{pages}{134112}
  (\bibinfo{year}{2017}).

\bibitem{lawley2019first}
\bibinfo{author}{Lawley, S.} \& \bibinfo{author}{Madrid, J.}
\newblock \bibinfo{title}{First passage time distribution of
  multiple impatient particles with reversible binding}.
\newblock \bibinfo{journal}{\emph{Journal of Chemical Physics}}
  \textbf{\bibinfo{volume}{150}}, \bibinfo{pages}{214113}
  (\bibinfo{year}{2019}).

\bibitem{rebola2019distinct}
\bibinfo{author}{Rebola, N.} \emph{et~al.}
\newblock \bibinfo{title}{Distinct nanoscale calcium channel
  and synaptic vesicle topographies contribute to the diversity of synaptic
  function}.
\newblock \bibinfo{journal}{\emph{Neuron}} \textbf{\bibinfo{volume}{104}},
  \bibinfo{pages}{693--710} (\bibinfo{year}{2019}).

\bibitem{wang2008synaptic}
\bibinfo{author}{Wang, L.-Y.}, \bibinfo{author}{Neher, E.} \&
  \bibinfo{author}{Taschenberger, H.}
\newblock \bibinfo{title}{Synaptic vesicles in mature calyx
  of held synapses sense higher nanodomain calcium concentrations during action
  potential-evoked glutamate release}.
\newblock \bibinfo{journal}{\emph{Journal of Neuroscience}}
  \textbf{\bibinfo{volume}{28}}, \bibinfo{pages}{14450--14458}
  (\bibinfo{year}{2008}).

\bibitem{vyleta2014loose}
\bibinfo{author}{Vyleta, N.~P.} \& \bibinfo{author}{Jonas, P.}
\newblock \bibinfo{title}{Loose coupling between ca2+
  channels and release sensors at a plastic hippocampal synapse}.
\newblock \bibinfo{journal}{\emph{Science}} \textbf{\bibinfo{volume}{343}},
  \bibinfo{pages}{665--670} (\bibinfo{year}{2014}).

\bibitem{yin2010}
\bibinfo{author}{Yin, G.} \& \bibinfo{author}{Zhu, C.}
\newblock \emph{\bibinfo{title}{Hybrid Switching Diffusions: Properties and
  Applications}} (\bibinfo{publisher}{Springer}, \bibinfo{address}{New York},
  \bibinfo{year}{2010}).

\bibitem{grebenkov2019}
\bibinfo{author}{Grebenkov, D.~S.}
\newblock \bibinfo{title}{A unifying approach to
  first-passage time distributions in diffusing diffusivity and switching
  diffusion models}.
\newblock \bibinfo{journal}{\emph{Journal of Physics A: Mathematical and Theoretical}}
  \textbf{\bibinfo{volume}{52}}, \bibinfo{pages}{174001}
  (\bibinfo{year}{2019}).

\bibitem{klafter2011first}
\bibinfo{author}{Klafter, J.} \& \bibinfo{author}{Sokolov, I.~M.}
\newblock \emph{\bibinfo{title}{First steps in random walks: from tools to
  applications}} (\bibinfo{publisher}{Oxford University Press},
  \bibinfo{year}{2011}).

\bibitem{shahrezaei2004consequences}
\bibinfo{author}{Shahrezaei, V.} \& \bibinfo{author}{Delaney, K.~R.}
\newblock \bibinfo{title}{Consequences of molecular-level Ca2+ 
channel and synaptic vesicle colocalization for the Ca2+ microdomain and
  neurotransmitter exocytosis: a monte carlo study}.
\newblock \bibinfo{journal}{\emph{Biophysical Journal}}
  \textbf{\bibinfo{volume}{87}}, \bibinfo{pages}{2352--2364}
  (\bibinfo{year}{2004}).

\bibitem{aracc2006close}
\bibinfo{author}{Ara{\c{c}}, D.} \emph{et~al.}
\newblock \bibinfo{title}{Close membrane-membrane proximity
  induced by Ca2+-dependent multivalent binding of synaptotagmin-1 to
  phospholipids}.
\newblock \bibinfo{journal}{\emph{Nature Structural \& Molecular Biology}}
  \textbf{\bibinfo{volume}{13}}, \bibinfo{pages}{209} (\bibinfo{year}{2006}).

\bibitem{allbritton1992range}
\bibinfo{author}{Allbritton, N.~L.}, \bibinfo{author}{Meyer, T.} \&
  \bibinfo{author}{Stryer, L.}
\newblock \bibinfo{title}{Range of messenger action of
  calcium ion and inositol 1, 4, 5-trisphosphate}.
\newblock \bibinfo{journal}{\emph{Science}} \textbf{\bibinfo{volume}{258}},
  \bibinfo{pages}{1812--1815} (\bibinfo{year}{1992}).

\bibitem{xu1997kinetic}
\bibinfo{author}{Xu, T.}, \bibinfo{author}{Naraghi, M.}, \bibinfo{author}{Kang,
  H.} \& \bibinfo{author}{Neher, E.}
\newblock \bibinfo{title}{Kinetic studies of Ca2+ binding and
  Ca2+ clearance in the cytosol of adrenal chromaffin cells}.
\newblock \bibinfo{journal}{\emph{Biophysical Journal}}
  \textbf{\bibinfo{volume}{73}}, \bibinfo{pages}{532--545}
  (\bibinfo{year}{1997}).

\bibitem{nagerl2000binding}
\bibinfo{author}{N{\"a}gerl, U.~V.}, \bibinfo{author}{Novo, D.},
  \bibinfo{author}{Mody, I.} \& \bibinfo{author}{Vergara, J.~L.}
\newblock \bibinfo{title}{Binding kinetics of calbindin-d 28k
  determined by flash photolysis of caged Ca2+}.
\newblock \bibinfo{journal}{\emph{Biophysical Journal}}
  \textbf{\bibinfo{volume}{79}}, \bibinfo{pages}{3009--3018}
  (\bibinfo{year}{2000}).

\bibitem{bucurenciu2008nanodomain}
\bibinfo{author}{Bucurenciu, I.}, \bibinfo{author}{Kulik, A.},
  \bibinfo{author}{Schwaller, B.}, \bibinfo{author}{Frotscher, M.} \&
  \bibinfo{author}{Jonas, P.}
\newblock \bibinfo{title}{Nanodomain coupling between Ca2+
  channels and Ca2+ sensors promotes fast and efficient transmitter release at
  a cortical gabaergic synapse}.
\newblock \bibinfo{journal}{\emph{Neuron}} \textbf{\bibinfo{volume}{57}},
  \bibinfo{pages}{536--545} (\bibinfo{year}{2008}).


\bibitem{dodge1967co}
\bibinfo{author}{Dodge Jr, F.~A.} \& \bibinfo{author}{Rahamimoff, R.}
\newblock \bibinfo{title}{Co-operative action of calcium ions in transmitter release at the neuromuscular junction}.
\newblock \bibinfo{journal}{\emph{Journal of Physiology}}
  \textbf{\bibinfo{volume}{193}}, \bibinfo{pages}{419--432} (\bibinfo{year}{1967}).

\bibitem{chapman2018ca2+}
\bibinfo{author}{Chapman, E.~R}
\newblock \bibinfo{title}{A Ca2+ sensor for exocytosis}.
\newblock \bibinfo{journal}{\emph{Trends in Neurosciences}}
  \textbf{\bibinfo{volume}{41}}, \bibinfo{pages}{327--330} (\bibinfo{year}{2018}).

\bibitem{schneggenburger2000intracellular}
\bibinfo{author}{Schneggenburger, R.} \& \bibinfo{author}{Neher, E.}
\newblock {\bibinfo{title}{Intracellular calcium dependence of transmitter release rates at a fast central synapse}}.
\newblock \bibinfo{journal}{\emph{Nature}}
  \textbf{\bibinfo{volume}{406}}, \bibinfo{pages}{889--893} (\bibinfo{year}{2000}).

\bibitem{bollmann2000calcium}
\bibinfo{author}{Bollmann, J.~H.}, \bibinfo{author}{Sakmann, B.} \& \bibinfo{author}{Borst, J. G. G.}
\newblock \bibinfo{title}{Calcium sensitivity of glutamate release in a calyx-type terminal}.
\newblock \bibinfo{journal}{\emph{Science}}
  \textbf{\bibinfo{volume}{289}}, \bibinfo{pages}{953--957} (\bibinfo{year}{2000}).



\bibitem{grebenkov2019reversible}
\bibinfo{author}{Grebenkov, D.~S.}
\newblock \bibinfo{title}{Reversible reactions controlled by
  surface diffusion on a sphere}.
\newblock \bibinfo{journal}{\emph{Journal of Chemical Physics}}
  \textbf{\bibinfo{volume}{151}}, \bibinfo{pages}{154103}
  (\bibinfo{year}{2019}).

\bibitem{grebenkov2003}
\bibinfo{author}{Grebenkov, D.~S.}, \bibinfo{author}{Filoche, M.} \&
  \bibinfo{author}{Sapoval, B.}
\newblock \bibinfo{title}{Spectral properties of the Brownian
  self-transport operator}.
\newblock \bibinfo{journal}{\emph{European Physical Journal B}}
  \textbf{\bibinfo{volume}{36}}, \bibinfo{pages}{221--231}
  (\bibinfo{year}{2003}).

\bibitem{grebenkov2019c}
\bibinfo{author}{Grebenkov, D.~S.}
\newblock \bibinfo{title}{Imperfect diffusion-controlled reactions}.
\newblock In \bibinfo{editor}{Lindenberg, K.}, \bibinfo{editor}{Metzler, R.} \&
  \bibinfo{editor}{Oshanin, G.} (eds.) \emph{\bibinfo{booktitle}{Chemical
  Kinetics: Beyond the Textbook}}, chap.~\bibinfo{chapter}{8},
  \bibinfo{pages}{191--219} (\bibinfo{publisher}{World Scientific},
  \bibinfo{year}{2019}).

\bibitem{helmchen1997calcium}
\bibinfo{author}{Helmchen, F.}, \bibinfo{Borst, J. G.}, \& \bibinfo{Sakmann, B.}
\newblock \bibinfo{title}{Calcium dynamics associated with a single action potential in a CNS presynaptic terminal}.
\newblock \bibinfo{journal}{\emph{Biophysical Journal}}
  \textbf{\bibinfo{volume}{72}}, \bibinfo{pages}{1458--1471} (\bibinfo{year}{1997}).

\bibitem{lou2005allosteric}
\bibinfo{author}{Lou, X.}, \bibinfo{author}{Scheuss, V.}, \& \bibinfo{author}{Schneggenburger, R.}
\newblock \bibinfo{title}{Allosteric modulation of the presynaptic Ca 2+ sensor for vesicle fusion}.
\newblock \bibinfo{journal}{\emph{Nature}}
  \textbf{\bibinfo{volume}{435}}, \bibinfo{pages}{497--501} (\bibinfo{year}{2005}).

\bibitem{agmon1990}
\bibinfo{author}{Agmon, N.} \& \bibinfo{author}{Szabo, A.}
\newblock \bibinfo{title}{Theory of reversible
  diffusion-influenced reactions}.
\newblock \bibinfo{journal}{\emph{Journal of Chemical Physics}}
  \textbf{\bibinfo{volume}{92}}, \bibinfo{pages}{5270} (\bibinfo{year}{1990}).

\bibitem{prustel2013}
\bibinfo{author}{Pr\"ustel, T.} \& \bibinfo{author}{Tachiya, M.}
\newblock \bibinfo{title}{Reversible diffusion-influenced
  reactions of an isolated pair on some two dimensional surfaces}.
\newblock \bibinfo{journal}{\emph{Journal of Chemical Physics}}
  \textbf{\bibinfo{volume}{139}}, \bibinfo{pages}{194103}
  (\bibinfo{year}{2013}).

\bibitem{redner2001}
\bibinfo{author}{Redner, S.}
\newblock \emph{\bibinfo{title}{A Guide to First Passage Processes}}
  (\bibinfo{publisher}{Cambridge University press},
  \bibinfo{address}{Cambridge}, \bibinfo{year}{2001}).

\bibitem{crank1975}
\bibinfo{author}{Crank, J.}
\newblock \emph{\bibinfo{title}{The Mathematics of Diffusion}}
  (\bibinfo{publisher}{Clarendon}, \bibinfo{address}{Oxford},
  \bibinfo{year}{1975}), \bibinfo{edition}{2} edn.

\bibitem{carslaw1959}
\bibinfo{author}{Carslaw, H.~S.} \& \bibinfo{author}{Jaeger, J.~C.}
\newblock \emph{\bibinfo{title}{Conduction of Heat in Solids}}
  (\bibinfo{publisher}{Clarendon}, \bibinfo{address}{Oxford},
  \bibinfo{year}{1975}), \bibinfo{edition}{2} edn.

\bibitem{grebenkov2018}
\bibinfo{author}{Grebenkov, D.~S.}, \bibinfo{author}{Metzler, R.} \&
  \bibinfo{author}{Oshanin, G.}
\newblock \bibinfo{title}{Strong defocusing of molecular
  reaction times results from an interplay of geometry and reaction control}.
\newblock \bibinfo{journal}{\emph{Communications Chemistry}}
  \textbf{\bibinfo{volume}{1}}, \bibinfo{pages}{96} (\bibinfo{year}{2018}).

\bibitem{godec2017}
\bibinfo{author}{Godec, A.} \& \bibinfo{author}{Metzler, R.}
\newblock \bibinfo{title}{First passage time statistics for
  two-channel diffusion}.
\newblock \bibinfo{journal}{\emph{Journal of Physics A: Mathematical and Theoretical}}
  \textbf{\bibinfo{volume}{50}}, \bibinfo{pages}{084001}
  (\bibinfo{year}{2017}).

\bibitem{shoup1982}
\bibinfo{author}{Shoup, D.} \& \bibinfo{author}{Szabo, A.}
\newblock \bibinfo{title}{Role of diffusion in ligand binding
  to macromolecules and cell-bound receptors}.
\newblock \bibinfo{journal}{\emph{Biophysical Journal}}
  \textbf{\bibinfo{volume}{40}}, \bibinfo{pages}{33} (\bibinfo{year}{1982}).

\bibitem{lauffenburger1993}
\bibinfo{author}{Lauffenburger, D.~A.} \& \bibinfo{author}{Linderman, J.}
\newblock \emph{\bibinfo{title}{Receptors: Models for Binding, Trafficking, and
  Signaling}} (\bibinfo{publisher}{Oxford University Press},
  \bibinfo{address}{Oxford}, \bibinfo{year}{1993}).

\bibitem{hodgkin1952quantitative}
\bibinfo{author}{Hodgkin, A.~L.} \& \bibinfo{author}{Huxley, A.~F.}
\newblock \bibinfo{title}{A quantitative description of
  membrane current and its application to conduction and excitation in nerve}.
\newblock \bibinfo{journal}{\emph{Journal of Physiology}}
  \textbf{\bibinfo{volume}{117}}, \bibinfo{pages}{500--544}
  (\bibinfo{year}{1952}).

\bibitem{kerr2008fast}
\bibinfo{author}{Kerr, R.~A.} \emph{et~al.}
\newblock \bibinfo{title}{Fast Monte Carlo simulation methods
  for biological reaction-diffusion systems in solution and on surfaces}.
\newblock \bibinfo{journal}{\emph{SIAM Journal on Scientific Computing}}
  \textbf{\bibinfo{volume}{30}}, \bibinfo{pages}{3126--3149}
  (\bibinfo{year}{2008}).






\bibitem{Yin10}			G. Yin and C. Zhu
				{\it Hybrid Switching Diffusions:
				Properties and Applications}
				(Springer, New York, 2010).

\bibitem{Yin10b}		G. Yin and C. Zhu,
				``Properties of solutions of stochastic differential equations with continuous-state-dependent switching'',
				J. Diff. Eq. {\bf 249}, 2409-2439 (2010).

\bibitem{Baran13}		N. A. Baran, G. Yin, and C. Zhu,
				``Feynman-Kac formula for switching diffusions: connections of systems of partial differential 
				equations and stochastic differential equations'',
				Adv. Diff. Eq. 2013:315 (2013).

\bibitem{Grebenkov19b}		D. S. Grebenkov, 
				``Time-averaged mean square displacement for switching diffusion'', 
				Phys. Rev. E {\bf 99}, 032133 (2019).



\bibitem{Freidlin}		M. Freidlin, 
				{\it Functional Integration and Partial Differential Equations}, 
				Annals of Mathematics Studies
				(Princeton University Press, Princeton, New Jersey, 1985).


\bibitem{Grebenkov06}		D. S. Grebenkov,  
				``Partially Reflected Brownian Motion: A Stochastic Approach to Transport Phenomena'',
				in ``Focus on Probability Theory'', Ed. L.~R.~Velle, pp. 135-169 (Nova Science Publishers, 2006).

\bibitem{Grebenkov07}		D. S. Grebenkov, 
				``Residence times and other functionals of reflected Brownian motion'',
				Phys. Rev. E {\bf 76}, 041139 (2007).

\bibitem{Singer08}		A. Singer, Z. Schuss, A. Osipov, and D. Holcman,
				``Partially Reflected Diffusion'',
				SIAM J. Appl. Math. {\bf 68}, 844 (2008).


\bibitem{Grebenkov18b}		D. S. Grebenkov, R. Metzler, and G. Oshanin, 
				``Strong defocusing of molecular reaction times results from an interplay of geometry and reaction control'', 
				Commun. Chem. {\bf 1}, 96 (2018).





\bibitem{Sheng12}		J. Sheng, L. He, H. Zheng, L. Xue, F. Luo, W. Shin, T. Sun, T. Kuner, D. T. Yue, and L.-G. Wu,
				``Calcium-channel number critically influences synaptic strength and plasticity at the active zone'',
				Nature Neurosci. {\bf 15}, 998-1006 (2012).









\end{thebibliography}

\section*{Acknowledgements}

This study was supported by the Centre National de la Recherche Scien- tifique, Fondation pour la Recherche Medicale (Equipe FRM), Agence Nationale de la Recherche (ANR-2010-BLANC-1411, ANR-13-BSV4-0016, ANR-17-CE16-0019, and ANR-17-CE16-0026), and Ile de France (Domaine d’Interet Majeur (DIM) MALINF: DIM120121). M.R. was supported by the Pasteur Paris University (PPU) doctoral program. The laboratory of D.D. is a member of the Bio-Psy Laboratory of Excellence. We would like to thank Yukihiro Nakamura for sharing simulated calcium transients, Jean-Baptiste Masson and Nelson Rebola for the comments on the manuscript and discussions. 

\section*{Author contributions statement}

M.R, D.D. and D.G. conceived the study, analysed results and wrote manuscript, D.G. made analytical derivations, M.R. performed simulations. 

\section*{Additional information}

The authors declare no competing interests.

\newpage

\setcounter{equation}{0}
\setcounter{section}{0}
\setcounter{figure}{0}
\setcounter{table}{0}
\setcounter{page}{1}
\makeatletter
\renewcommand{\theequation}{S\arabic{equation}}
\renewcommand{\thefigure}{S\arabic{figure}}
\renewcommand{\thesection}{\Roman{section}}

{\Large \bf Supplementary Information}

\section{Formal definition of the switching diffusion model}
\label{sec:Amodel}

We reproduce here a formal definition of the $(M+1)$-state switching
diffusion model following Ref. \cite{Yin10}.  We consider a
two-component process $(\X_t,\nu_t)$, in which $\X_t$ is the diffusion
process in $\R^3$, and $\nu_t$ is the pure jump process with the
states at $\{0,1,\ldots,M\}$.  When there is no boundary, the process
is defined by a standard stochastic equation
\begin{equation}
d\X_t = \sqrt{2D_{\nu_t}}\, I \, d\W_t, \qquad (\X_0,\nu_0) = (\x,i),
\end{equation}
where $\W_t$ is the standard Wiener process in $\R^3$, $I$ is the
identity matrix, and $D_i$ is the diffusion coefficient at the state
$i$.  The jump process is defined for any $i \ne j$ by
\begin{equation}
\P\{ \nu_{t+dt} = j ~|~ \nu_t = i,\, \X_s, \nu_s, s\leq t\} = k_{ij} dt + o(dt),
\end{equation}
where $k_{ij}$ is the rate of transition from the state $i$ to the
state $j$.  The propagator $p(\x,i,t|\x_0,i_0,0)$ is the probability
density for the process to be in (the vicinity of) the point $\x$ in
the state $i$ at time $t$ when stated from the point $\x_0$ in the
state $i_0$.  The propagator satisfies $(M+1)$ coupled forward
Fokker-Planck equations
\begin{align}  
& \partial_t p(\x,i,t|\x_0,i_0,0) = D_i \Delta p(\x,i,t|\x_0,i_0,0)  + \sum\limits_{j=0}^M  \bigl[k_{ji} p(\x,j,t|\x_0,i_0,0) - k_{ij} p(\x,i,t|\x_0,i_0,0)\bigr], 
\end{align}
subject to the initial condition $p(\x,i,0|\x_0,i_0,0) =
\delta_{i,i_0} \, \delta(\x-\x_0)$.  Some properties of the propagator
were discussed in \cite{Yin10b,Baran13,Grebenkov19b} (see also the
references therein).

In turn, for a given smooth function $f$, the expectation of a
functional $f(\X_t,\nu_t)$ given that the process has started at $\x$
and $i$,
\begin{equation}
u(\x,i,t) = \E\{ f(\X_t,\nu_t) ~|~ X_0 = \x, \nu_0 = i\},
\end{equation}
satisfies the $(M+1)$ coupled backward Fokker-Planck (or Kolmogorov)
equations for each $i$,
\begin{equation}  \label{eq:backward}
\partial_t u(\x,i,t) = D_i \Delta u(\x,i,t) + \sum\limits_{j=0}^M k_{ij} \bigl( u(\x,j,t) - u(\x,i,t)\bigr), 
\end{equation}
subject to the initial condition $u(\x,i,0) = f(\x,i)$ (strictly
speaking, this is a terminal condition but as the rates $k_{ij}$ do
not depend on time, one can recast it as the initial condition).

In the presence of a (partially) reflecting boundary, the diffusion
component of the process is modified in a standard way (via the
Skorokhod equation) \cite{Freidlin,Grebenkov06,Grebenkov07,Singer08},
whereas the forward and backward Fokker-Planck equations need to be
completed by the associated boundary conditions, see
\cite{Yin10,Yin10b,Baran13}.  Setting $f = 1$, one can interpret
$u(\x,i,t)$ as the probability for a particle started at $\x$ in the
state $i$ to survive up to time $t$.

\section{General analytical solution}
\label{sec:AInversion}

In this section, we present the derivation of the analytical solution
for a general case with $M$ buffers.  Two particular cases (without
buffer and with one buffer) will be detailed in Sections
\ref{sec:AexampleM0} and \ref{sec:AexampleM1}.

\subsection{Survival probabilities}

We aim to find the survival probabilities $S_i(t,\x)$ satisfying
Eqs. (\ref{eq:backward}) with $f = 1$ inside the domain 
\begin{equation}
\Omega = \{ \x \in \R^3 ~:~ \rho < |\x| < R\} 
\end{equation}
between two concentric spheres of radii $\rho$ and $R$.  The rotation
symmetry of this domain implies that $S_i(t,\x)$ depend only on the
radial coordinate $r = |\x|$ so that we can drop the dependence on
angular coordinates and write $S_i(t,r)$.  Equations
(\ref{eq:backward}) are subject to the initial condition
\begin{equation}
S_i(t=0,r) = 1 ,
\end{equation}
and have to be completed by boundary conditions (see the main text)
\begin{subequations}
\begin{eqnarray}
\rho \bigl(\partial_r S_i(t,r)\bigr)_{r=\rho} &=& \mu_i \, S_0(t,\rho), \\
\bigl(\partial_r S_i(t,r)\bigr)_{r=R} &=& 0, 
\end{eqnarray}
\end{subequations}
at the inner and outer spheres, respectively, where
\begin{equation}
\label{eq:hkon}
\mu_0 = \mu = \frac{\kon}{4\pi \rho D_0 N_A} ,  \qquad \mu_i = 0 \quad (i=1,\ldots,M)
\end{equation}
are dimensionless reactivities, with $N_A$ being the Avogadro number,
and $\kon$ the on-rate binding constant.

Introducing the Laplace-transformed survival probabilities (denoted by
tilde),
\begin{equation}
\tilde S_i(p,r) = \int\limits_0^\infty dt \, e^{-pt} \, S_i(t,r),
\end{equation}
one can rewrite the above equations as
\begin{subequations}  \label{eq:tildeS}   
\begin{align}
(p + k_i - D_i \Delta) \tilde{S}_i - \sum\limits_{j=0}^M k_{ij} \tilde S_j &= 1  \quad (\rho < r < R), \\
\partial_r \tilde{S}_i &=  0 \quad (r=R) , \\
\mu_i \tilde{S}_i - \rho \, \partial_r \tilde{S}_i &= 0 \quad (r=\rho) ,
\end{align}
\end{subequations}
where $\Delta = \partial_r^2 + (2/r) \partial_r$ is the radial part of
the Laplace operator, and
\begin{equation}
k_i = \sum\limits_{j=0}^M k_{ij} .
\end{equation}
As the rate $k_{ii}$ is undefined, we set $k_{ii} = 0$ for convenience
of notations.

We search the Laplace-transformed probabilities in the form
\begin{equation}
\label{eq:tildeSi_form}
\tilde S_i(p,r) = a_i + \sum\limits_{j=0}^{M} b_{ij}\,  v(\delta_j,r) ,
\end{equation}
where $a_i$ and $b_{ij}$ are unknown coefficients, and
\begin{equation}
\label{eq:vr}
v(\delta, r) = \frac{\rho}{r} \biggl( \sinh(\delta (R-r)/\rho) - (1+\beta) \delta \cosh(\delta (R-r)/\rho)\biggr) ,
\end{equation}
with 
\begin{equation}
\beta = (R-\rho)/\rho , 
\end{equation}
and $\delta_j$ are unknown factors.  In fact, the function
$v(\delta,r)$ is a linear combination of two independent solutions
$e^{\delta r}/r$ and $e^{-\delta r}/r$ of the equation $\Delta u -
\delta^2 u = 0$, and the chosen form (\ref{eq:vr}) ensures the Neumann
boundary condition at the outer sphere for any $\delta$:
\begin{equation}
\bigl(\partial_r v(\delta,r)\bigr)_{r = R} = 0 . 
\end{equation}
Substituting Eq. (\ref{eq:tildeSi_form}) into Eq. (\ref{eq:tildeS}),
we get for $i=0,\ldots,M$
\begin{align} 
& (p+k_i) \biggl(a_i + \sum\limits_{j=0}^M b_{ij}\, v(\delta_j,r) \biggr) 
 - \frac{D_i}{\rho^2} \sum\limits_{j=0}^M b_{ij}  \delta_j^2 \, v(\delta_j,r)  - \sum\limits_{\ell=0}^M k_{i\ell} \biggl(a_\ell + 
\sum\limits_{j=0}^M b_{\ell j} \, v(\delta_j,r) \biggr) = 1  
\end{align}
Each of these $M+1$ functional relations must be satisfied for any $r
\in (\rho,R)$ that implies $M+2$ relations on coefficients for each
$i=0,\ldots,M$:
\begin{equation}
\label{eq:LinEq1}
(p+k_i) a_i - \sum\limits_{\ell=0}^M k_{i\ell} a_\ell = 1
\end{equation}
and
\begin{equation}
\label{eq:LinEq2}
\bigl(p+k_i - (D_i/\rho^2) \delta_j^2\bigr) b_{ij} - \sum\limits_{\ell=0}^M k_{i\ell} b_{\ell j} = 0 .
\end{equation}

The first set (\ref{eq:LinEq1}) of $M+1$ linear equations on $a_i$ is
uncoupled from the rest and can be solved separately.  Inverting the
underlying matrix,
\begin{equation}
\label{eq:W}
W = \left(\begin{array}{ccccc} 
\gamma_{0} & - k_{01}   & - k_{02}   & \cdots & - k_{0M} \\
- k_{10}   & \gamma_{1} &     0      & \cdots &     0    \\
- k_{20}   &     0      & \gamma_{2} & \cdots &     0    \\
   \cdots  &    \cdots  &   \cdots   & \cdots & \cdots   \\
- k_{M0}   &     0      &     0      & \cdots & \gamma_{M} \\  \end{array} \right)
\end{equation}
(with $\gamma_i = p+k_i$) and applying to the vector
$(1,1,\ldots,1)^\dagger$, one gets
\begin{equation}
\label{eq:ai}
a_0 = \biggl(1 + \sum\limits_{i=1}^M \frac{k_{0i}}{p + k_{i0}} \biggr) 
\biggl(p + k_0 - \sum\limits_{i=1}^M \frac{k_{0i} k_{i0}}{p+k_{i0}}\biggr)^{-1} = \frac{1}{p}.  
\end{equation}
The other $a_i$ can also be found but their contribution will be
canceled by $\mu_i = 0$ for $i > 0$.

Next, we can treat Eqs. (\ref{eq:LinEq2}) as a set of linear equations
on $b_{ij}$, in which $\delta_j$ are some parameters.  One can note
that, for each $j$, there are $M+1$ equations whose form does not
depend on $j$.  In other words, we can decouple these equations into
$M+1$ blocks, each having $M+1$ equations.  Let us write $\delta$
instead of $\delta_j$ for one block.  The equations in each block are
homogeneous, so that there is either none, or infinitely many
solutions.  For the existence of solutions, the determinant of the
underlying matrix in front of coefficients $b_{ij}$ should be zero.
This matrix has precisely the same form as $W$ in Eq. (\ref{eq:W}),
but with $\gamma_i = p + k_i - (D_i/\rho^2) \delta^2$.  The
determinant of this matrix as a function of $z = \delta^2$ is the
polynomial of degree $(M+1)$
\begin{equation}
\label{eq:F_delta}
H(z) = \gamma_1 \cdots \gamma_M \biggl(\gamma_0 - \sum\limits_{i=1}^M \frac{k_{0i} k_{i0}}{\gamma_i}\biggr).
\end{equation}
The $M+1$ zeros of this polynomial, $z_i$, determine the unknown
$\delta_i$: $\delta_i = \sqrt{z_i}$ (here one can use either of two
values $\pm \sqrt{z_i}$, the final results remaining unchanged).

For each $j$, the set (\ref{eq:LinEq2}) of equations on $b_{ij}$ has
infinitely many solutions.  One can express $b_{ij}$ (for $i =
1,\ldots,M$) in terms of $b_{0j}$ as
\begin{equation}
\label{eq:bij}
b_{ij} = \frac{k_{i0}}{p + k_i - (D_i/\rho^2) \delta_j^2} \, b_{0j} .
\end{equation}

The remaining $M+1$ unknowns $b_{0j}$ are determined by the $(M+1)$
boundary conditions at the inner sphere:
\begin{equation}
\bigl(\mu_i \tilde S_i(p,r) - \rho\, \partial_r \tilde S_i(p,r)\bigr)_{r=\rho} = 0  \quad (i=0,\ldots,M)
\end{equation}
that implies $(M+1)$ linear relations
\begin{equation}
\label{eq:LinEq4}
\sum\limits_{j=0}^M b_{ij} c_{ij} = a_i \mu_i \qquad (i = 0,\ldots,M),
\end{equation}
where
\begin{align}  \nonumber
c_{ij} & = \biggl( \rho \bigl(\partial_r v(\delta_j,r)\bigr)_{r=\rho} - \mu_i v(\delta_j,\rho)\biggr) \\ 
& = \sinh(\beta \delta_j) \bigl((1+\beta)\delta_j^2 - 1 - \mu_i\bigr)  
 + \delta_j \cosh(\beta \delta_j) \bigl(\beta + \mu_i(1+\beta)\bigr). 
 \label{eq:cij} 
\end{align}
Substituting Eqs. (\ref{eq:bij}) into these relations, one gets $M+1$
linear equations on the remaining $M+1$ unknowns $b_{0j}$:
\begin{equation}
\label{eq:LinEq4b}
\sum\limits_{j=0}^{M} C_{ij} b_{0j} = a_i \mu_i \qquad (i=0,\ldots,M),
\end{equation}
with
\begin{equation}
\label{eq:C}
C_{ij} = c_{ij} \times \begin{cases}  1  \hskip 32mm (i = 0) , \cr 
\displaystyle \frac{k_{i0}}{p + k_i - (D_i/\rho^2) \delta_j^2}  \quad (i > 0) . \end{cases}
\end{equation}
Inverting the matrix $C$, one obtains $b_{0j}$ and thus fully
determines $\tilde S_i(p,r)$.  Given that $\mu_i = 0$ for $i > 0$,
$b_{0j}$ can formally be written as
\begin{equation}
b_{0j} = \frac{\mu \, f_{0j}(p)}{p \, f(p)} ,
\end{equation}
with
\begin{equation}
\label{eq:fp}
f(p) = \det(C), \qquad f_{ij}(p) = (-1)^{i+j} {\mathcal C}_{ij} ,
\end{equation}
where ${\mathcal C}_{ij}$ is the $(i,j)$ minor of $C$, i.e., the
determinant of the $M \times M$ matrix that results from deleting row
$i$ and column $j$ of $C$.  We get thus
\begin{equation}
\label{eq:tildeS0_final}
\tilde{S}_0(p,r) = \frac{1}{p} \biggl(1 + \frac{w(p,r)}{f(p)} \biggr),
\end{equation}
where 
\begin{equation}
w(p,r) = \mu \sum\limits_{j=0}^M f_{0j}(p) \, v(\delta_j,r) .
\end{equation}
This is the exact analytic solution of the problem in the Laplace
domain.  In order to get the solution in time domain, one needs to
compute the poles of $\tilde S_0(p,r)$ which are given by zeros of the
function $f(p)$ considered in the complex plane ($p \in \C$).

The survival probability $\tilde{S}_0(p,r)$ also determines the
probability density of the first binding time, $\tilde{\psi}_1(p,r) =
1 - p \tilde{S}_0(p,r)$, from which
\begin{equation}  \label{eq:tilde_psi1A}
\tilde{\psi}_1(p,r) = -  \frac{w(p,r)}{f(p)} \,.
\end{equation}
In the general case $k_{i0} > 0$ (i.e., when buffers cannot bind
calcium ions forever), one can show that $\tilde{\psi}_1(0,r) = 1$
that corresponds to the normalization of the probability density
$\psi_1(t,r)$ (we omit the related asymptotic analysis of the minors
$f_{ij}(p)$ and of $f(p)$ as $p\to 0$; see the example for one buffer
in Sec. \ref{sec:AexampleM1}).  As a consequence, $p = 0$ is not a
pole of $\tilde{S}_0(p,r)$, and $S_0(t,r)$ vanishes in the long time
limit.  In turn, if $k_{i0} = 0$ for some $i$, the calcium ions can be
trapped forever by that buffer, and $\tilde{S}_0(t,r)$ reaches a
nonzero limit (the fraction of such trapped ions).  In this specific
case, $\tilde{\psi_1}(0,r) < 1$, i.e., the normalization of
$\psi_1(t,r)$ does not hold.  In practice, even if $k_{i0}$ is very
small, it is nonzero, and this pathologic situation does not occur.
Note also that $\tilde{S}_0(p,r)$ determines the moments of the first
binding times; in particular, the mean time is simply
\begin{equation}
\langle \t \rangle = \int\limits_0^\infty dt \, t \, \psi_1(t,r) = \int\limits_0^\infty dt \, S_0(t,r) = \tilde{S}_0(0,r),
\end{equation}
where we integrated by parts and used that $\psi_1(t,r) = -\partial_t
S_0(t,r)$ and $S_0(\infty,r) = 0$.

\subsection{Occupancy probability}

As discussed in the Methods Section, the probability density of the
first binding times determines the occupancy probability $P(t,r)$ in
the Laplace domain as
\begin{equation} \label{eq:tildePPA}
\tilde P(p,r) = \tilde{\psi}_1(p,r) \, \tilde{Q}(p), 
\end{equation}
where 
\begin{equation}  \label{eq:tildeQ}
\tilde{Q}(p) = \biggl(p + \koff (1 - \tilde{\psi}(p,\rho))\biggr)^{-1} .
\end{equation}
Substituting Eq. (\ref{eq:tilde_psi1A}) into this equation yields
\begin{equation}
\tilde{Q}(p) = \biggl(p + \koff + p \koff \sum\limits_{j=0}^M b_{0j} \, v(\delta_j, \rho)\biggr)^{-1} \,.
\end{equation}
Next, substituting this expression and Eq. (\ref{eq:b00b01}) into
Eq. (\ref{eq:tildePPA}), we get explicitly
\begin{equation}
\label{eq:tildeP_final}
\tilde{P}(p,r) = - \frac{w(p,r)}{F(p)} \,, 
\end{equation}
with
\begin{equation}
\label{eq:Fp}
F(p) = (p+\koff)f(p) + \koff \, w(p,\rho).
\end{equation}
The poles of $\tilde{P}(p,r)$ are given by zeros of the function $F(p)$:
\begin{equation}
F(p_n) = 0  \qquad (n= 0,1,\ldots) .
\end{equation}
One can invert the Laplace transform by using the residue theorem.  In
particular, if the poles are simple, one gets
\begin{equation}
\label{eq:Pt_general}
P(t,r) = \sum\limits_{n=0}^\infty \bar{b}_n \, w(p_n,r) \, \exp(p_n t) ,
\end{equation}
where
\begin{equation}
\bar{b}_n = - \frac{1}{\lim\limits_{p\to p_n} \partial_p F(p)} \, ,
\end{equation}
in which the derivative can be computed by using 
\begin{equation}
\frac{\partial}{\partial p} v(\delta,r) = - \biggl(\cosh(\beta \delta) + 
\beta(1+\beta)\delta \sinh(\beta\delta)\biggr) \frac{\partial \delta}{\partial p} 
\end{equation}
and
\begin{align}  
& \frac{\partial}{\partial p} c_{ij} = \frac{\partial \delta_j}{\partial p}  
\biggl\{ \cosh(\beta \delta_j) \bigl(\mu_i + \beta (1+\beta)\delta_j^2\bigr) 
 + \delta_j \sinh(\beta \delta_j) \bigl(\mu_i \beta(1+\beta) + (\beta^2 + 2\beta+2)\bigr) \biggr\}  .
 \label{eq:dc_ij}
\end{align}
It can checked that $p_0 = 0$ whereas the other poles are strictly
negative: $p_n < 0$.  As a consequence, as $t\to \infty$, the
probability $P(t,r)$ approaches a stationary value $P_\infty$, which
is independent of the starting point $r$ and given by the residue at
$p_0 = 0$.  Summarizing these results, the occupancy probability takes
the form
\begin{equation}
\label{eq:Pt_general2}
P(t,r) = P_\infty + \sum\limits_{n=1}^\infty \exp(p_n t) \sum\limits_{j=0}^M b_n^j \, v(\delta_j(p_n),r) \,  ,
\end{equation}
where
\begin{equation}
b_n^j = \mu \, f_{0j}(p_n) \, \bar{b}_n.  
\end{equation}
Setting 
\begin{align*}
& \alpha_n = \rho \sqrt{- p_n/D_0},  \quad 
\alpha_n^{(j)} = -i \delta_j(p_n), \quad
b^{(j)}_n = i b^j_n, 
\end{align*}
one can rewrite the occupancy probability in a more conventional form:
\begin{equation}  \label{eq:Pt_general3}
P(t,r) = P_\infty + \sum\limits_{n=1}^\infty \exp\bigl(- \alpha_n^2 D_0 t/\rho^2\bigr) \sum\limits_{j=0}^M b_n^{(j)} \, u(\alpha_n^{(j)},r) ,
\end{equation}
where
\begin{equation}
\label{eq:urA}
u(\delta,r) = \frac{\rho}{r} \biggl(\sin(\delta (R-r)/\rho) - (1+\beta)\delta \cos(\delta (R-r)/\rho)\biggr) .
\end{equation}

We note that the functions $\tilde{S}_0(p,r)$ and $\tilde{P}(p,r)$
involve complicated combinations of roots (e.g., square roots, see
below) emerging from the zeros of the polynomial $H(z)$ in
Eq. (\ref{eq:F_delta}).  As a consequence, the use of the residue
theorem for evaluating the inverse Laplace transform of these
functions is not straightforward as one needs to introduce cuts in the
complex plane to properly deal with such multivariate functions.  In
addition, the application of Eq. (\ref{eq:Pt_general}) relies on the
assumption of simple poles.  In this paper, we do not provide rigorous
mathematical analysis of both statements.  In turn, we have checked
the correctness and the accuracy of the derived formulas in time
domain by comparison with the numerical inversion of the Laplace
transform (not shown).

In summary, the analytic solution requires three numerical steps: (i)
computation of $\delta_j^2$ as the zeros of Eq. (\ref{eq:F_delta});
(ii) inversion of the matrix $C$ in Eq. (\ref{eq:C}), from which
$f_{0j}(p)$, $f(p)$ and thus $b_{0j}$ are found; and (iii) finding the
zeros of $f(p)$ (for getting $S_0(t,r)$) or of $F(p)$ (for getting
$P(p,r)$) for the inversion of the Laplace transform.  We emphasize
that $\delta_j$ and $b_{0j}$ depend on $p$, i.e. one needs to perform
the first two steps for all values of $p$ at which $\tilde S_0(p,r)$
has to be found.  In practice, the number of buffers, $M$, is not
large so that these numerical steps can be done very rapidly and with
any accuracy.  We will discuss the cases $M = 0$ (Sec.
\ref{sec:AexampleM0}) and $M = 1$ (Sec. \ref{sec:AexampleM1}), for
which (some of) these steps can be done analytically.

\subsection{Steady-state limit $P_\infty$}

As time $t$ goes to infinity, the occupancy probability $P(t,r)$ from
Eq. (\ref{eq:Pt_general3}) approaches the steady-state limit
$P_\infty$, which is determined by the residue of $\tilde P(p,r)$ at
the pole $p = 0$.  Even though all the formulas determining $\tilde
P(p,r)$ are given, the computation of this residue is technically
involved, see the related analysis below for the particular cases of
no buffer and one buffer.  For this reason, we prefer to rely here on
qualitative physical arguments that allow us to get the exact form of
$P_\infty$ without tedious computations.  

In the steady-state, the system reaches an equilibrium between the
free state, the buffer-bound states, and the sensor-bound state.
Moreover, as the binding/unbinding kinetics on the sensor occurs only
through the free state, one can separate the kinetics with the
sensor and the kinetics with the buffers.  The equilibrium kinetics
with the sensor can be understood as a two-state switching model,
governed by two exchange rates: $\koff$ describes the transition from
the sensor-bound state to the free state, whereas an effective rate
$k_{0,s} = \kon \, c_0$ characterizes the opposite transition, where
$c_0$ is the equilibrated (homogeneous) concentration of calcium ions.
If $p_0$ is the equilibrium fraction of calcium ions in the free
state, then the conventional concentration (in M = mol/liter) reads
$c_0 = p_0/(N_A V)$, where $V = 4\pi (R^3 - \rho^3)/3$ is the volume
of the active zone.  In this setting, the occupancy probability (i.e.,
the probability of finding the calcium ion bound to the sensor) is
simply $P_\infty = k_{0,s}/(k_{0,s} + \koff)$ or, equivalently,
\begin{equation}
P_\infty = \frac{1}{1 + \koff \frac{N_A V}{\kon \, p_0}} \,.
\end{equation}

The fraction $p_0$ of calcium ions in the free state can be determined
from the equilibrium between the free state and buffer-bound states.
For this purpose, we only consider the dynamics of the $(M+1)$-state
switching model governed by the transition matrix $W$ from
Eq. (\ref{eq:W}) with $\gamma_i = k_i$ (i.e., at $p = 0$).  The
steady-state distribution is determined by the eigenvector of
$W^\dagger$ that corresponds to the eigenvalue $0$: $(1,
k_{01}/k_{10}, k_{02}/k_{20}, \ldots, k_{0M}/k_{M0})^\dagger$.  After
normalization to $1$, the probability of finding the calcium ion in
the free state (i.e., the fraction of calcium ions in this state) is
\begin{equation}
p_0 = \biggl(1 + \sum\limits_{j=1}^M \frac{k_{0j}}{k_{j0}}\biggr)^{-1} .
\end{equation}
We get therefore
\begin{equation}  \label{eq:PinfA}
P_\infty = \biggl(1 + \koff \frac{N_A V}{\kon} \biggl(1 + \sum\limits_{j=1}^M \frac{k_{0j}}{k_{j0}}\biggr) \biggr)^{-1} \,.
\end{equation}
The same expression for $P_\infty$ is retrieved for cases $M=0$
(Sec. \ref{sec:AexampleM0}) and $M=1$ (Sec. \ref{sec:AexampleM1}) from
the rigorous computation of the residue.

\section{No buffer case}
\label{sec:AexampleM0}

The survival probability for no buffer case is well known (see
\cite{Grebenkov18b} and references therein).  For illustrative
purposes, we retrieve this survival probability from our general
approach.  This step is also needed for finding the occupancy
probability $P(t,r)$.

When there is no buffer ($M = 0$), Eq. (\ref{eq:F_delta}) is reduced
to $H = \gamma_0 = p - (D_0/\rho^2)\delta^2 = 0$, from which $\delta_0
= \rho\sqrt{p/D_0}$.  The matrix $C$ consists of a single element
$C_{00} = c_{00}$, from which $f_{00}(p) = 1$, $f(p) = c_{00}$, and
thus $b_{00} = \mu/(p c_{00})$.  The Laplace-transformed survival
probability becomes then
\begin{equation}
\label{eq:tildeS_nobuf}
\tilde S_0(p,r) = \frac{1}{p} + \frac{\mu \, v(\delta_0,r)}{p \, c_{00}(p)} ,
\end{equation}
where $c_{00}(p)$ is given by Eq. (\ref{eq:cij}).  Setting $\delta_0 =
i\hat{\alpha}$, one can rewrite the equation $f(p) = 0$ on the poles of
$\tilde S_0(p,r)$ in a trigonometric form
\begin{equation}
\label{eq:alpha0_M0}
\sin(\hat{\alpha} \beta) = \frac{(\beta+(1+\beta)\mu)\, \hat{\alpha}}{1+ \mu + (1+\beta) \hat{\alpha}^2} \cos(\hat{\alpha} \beta),
\end{equation}
which has infinitely many nonnegative solutions denoted as
$\hat{\alpha}_n$, enumerated by $n = 0,1,2,\ldots$ (we use hat symbol
here to distinguish the quantities determining $S_0(t,r)$ from similar
quantities determining $P(t,r)$ below).  The poles are $\hat{p}_n = -
D_0 \hat{\alpha}_n^2/\rho^2$.  Note that the pole corresponding to
$\hat{\alpha}_0 = 0$ provides the contribution $-1/p$ that precisely
compensates the term $a_0 = 1/p$, and thus it will be excluded.  The
inverse Laplace transform is then obtained by the residue theorem:
\begin{equation}
\label{eq:A_S_final}
S_0(t,r) = \sum\limits_{n=1}^\infty \hat{b}_n  \, u(\hat{\alpha}_n,r) \, \exp\bigl(- \hat{\alpha}_n^2 D_0 t/\rho^2\bigr) ,
\end{equation}
where
\begin{align}  
\hat{b}_n & = \frac{2\mu}{\hat{\alpha}_n} \biggl(\cos(\hat{\alpha}_n\beta)\bigl[\mu - \beta(\beta+1) \hat{\alpha}_n^2\bigr] 
- \hat{\alpha}_n \sin(\hat{\alpha}_n\beta)\bigl[\beta(\beta+1)\mu + (\beta^2+2\beta+2)\bigr] \biggr)^{-1}
\label{eq:bn0_M0}
\end{align}
and $u(\delta,r)$ is given by Eq. (\ref{eq:urA}).  The derivative with
respect to $t$ yields the probability density of the first-binding
time
\begin{equation}
\label{eq:Apsi1_final}
\psi_1(t,r) = \frac{D_0}{\rho^2} \sum\limits_{n=1}^\infty  \hat{\alpha}_n^2 \, 
\hat{b}_n \, u(\hat{\alpha}_n,r)\, \exp\bigl(- \hat{\alpha}_n^2 D_0t/\rho^2\bigr) .
\end{equation}

\subsection*{Occupancy probability $P(t,r)$}

The computation of the probability $P(t,r)$ follows the same lines.
Setting $\delta_0 = i\alpha$ in Eq. (\ref{eq:Fp}), one gets
\begin{equation}
\label{eq:Falpha}
\begin{split}
F & = \frac{iD_0}{\rho^2} \biggl\{ \sin(\alpha\beta) \biggl((\alpha^2 - \lambda)((1+\beta)\alpha^2 + 1 + \mu) + \lambda \mu\biggr) 
 - \alpha \cos(\alpha \beta) \biggl((\alpha^2 - \lambda)(\beta + \mu(1+\beta)) + \lambda \mu (1+\beta) \biggr) \biggr\} , 
\end{split}
\end{equation}
from which the equation on $\alpha$ reads
\begin{equation}
\label{eq:alpha_M0A}
\sin(\alpha \beta) = \frac{\bigl[\alpha^2(\beta+\mu(1+\beta)) - \lambda \beta\bigr] \alpha \cos(\alpha \beta)}
{\alpha^4(1+\beta) + \alpha^2(1+\mu - \lambda(1+\beta)) - \lambda}  ,
\end{equation}
where $\lambda = \koff \rho^2/D_0$.  This equation has infinitely many
positive zeros that we denote as $\alpha_n$, with $n = 1,2,\ldots$
(the zero $\alpha_0 = 0$ will be considered separately).  These zeros
determine the poles: $p_n = - D_0 \alpha_n^2/\rho^2$.  Since $w(p_n,r)
= i \mu u_{\alpha_n}(r)$ with $u(\delta,r)$ given by
Eq. (\ref{eq:urA}), we obtain by the residue theorem
\begin{equation}
\label{eq:APt}
P(t,r) = \frac{1}{1 + \lambda \frac{(\beta+1)^3 - 1}{3\mu}} + \sum\limits_{n=1}^\infty b_n \, u(\alpha_n,r) \, e^{-\alpha_n^2 D_0 t/\rho^2} \,,  \\
\end{equation}
where the first term comes from the residue at $p = 0$, and
\begin{equation}
b_n = - \frac{i\mu}{\lim\limits_{p\to p_n} \bigl(\partial_p F(p)\bigr)}
= \frac{i\mu}{\frac{\rho^2}{2D_0 \alpha_n} \bigl(\partial_\alpha F(\alpha)\bigr)_{\alpha = \alpha_n}} . 
\end{equation}
Recalling the definition of dimensionless parameters $\lambda$, $\mu$
and $\beta$, one easily checks that the first term in
Eq. (\ref{eq:APt}) coincides with the steady-state limit $P_\infty$
in Eq. (\ref{eq:PinfA}).

Taking the derivative of Eq. (\ref{eq:Falpha}) with respect to
$\alpha$, one gets an explicit formula for $b_n$:
\begin{equation}
\label{eq:bn_M0A}
b_n = \frac{2 \mu}{\sin(\alpha_n \beta) \bigl(\alpha_n^2 w_1 + w_2 \bigr) 
 + \alpha_n \cos(\alpha_n\beta) \bigl(\alpha_n^2 w_3 + w_4 \bigr)} , 
\end{equation}
with
\begin{equation}
\begin{split}
w_1 & = 4(1+\beta) + \beta(\beta+\mu(1+\beta)) , \\
w_2 & = 2(1+\mu-\lambda(1+\beta)) - \lambda \beta^2 , \\
w_3 & = \beta(1+\beta) ,\\
w_4 & = \beta(1 + \mu - \lambda(1+\beta)) - 3(\beta + \mu(1+\beta))  .\\
\end{split}
\end{equation}

\subsection*{Limiting cases}

In the limit $\koff = 0$ (or $\lambda = 0$), there is no desorption
event, and $\tilde{Q}(p) = 1/p$ according to Eq. (\ref{eq:tildeQ}).
In this case, 
\begin{equation*}
\tilde{P}(p,r) = \frac{\tilde{\psi}_1(p,r)}{p} = \frac{1 - \tilde{S}_0(p,r)}{p} \, ,
\end{equation*}
and thus $P(t,r) = 1 - S_0(t,r)$, as expected.  One can also check
that the solutions $\alpha_n$ coincide with $\hat{\alpha}_n$.

In turn, in the limit of perfectly adsorbing sensor (i.e., with
infinitely fast binding kinetics: $\kon = \mu = \infty$),
Eq. (\ref{eq:alpha0_M0}) is reduced to
\begin{equation}
\sin(\hat{\alpha}_n \beta) = (1+\beta) \hat{\alpha}_n \cos(\hat{\alpha}_n \beta) ,
\end{equation}
and the survival probability becomes
\begin{equation}
\begin{split}
& S_0(t,r) = \frac{2\rho}{r} \sum\limits_{n=1}^\infty \exp\bigl(- \hat{\alpha}_n^2 D_0 t/\rho^2\bigr)  
 \times \frac{\sin\bigl(\hat{\alpha}_n \frac{R-r}{\rho}\bigr) - (1+\beta) \hat{\alpha}_n \cos\bigl(\hat{\alpha}_n \frac{R-r}{\rho}\bigr)}
{\hat{\alpha}_n \bigl(\cos (\hat{\alpha}_n \beta) - \beta(1+\beta)\hat{\alpha}_n \sin(\hat{\alpha}_n \beta)\bigr)} . \\
\end{split}
\end{equation}
The probability density $\psi_1(t,r) = - \partial_t S_0(t,r)$ is
obtained by taking the derivative with respect to $t$.  Note that in
this limit, the unbinding events are effectively suppressed as a
particle that unbinds from such a sensor immediately re-binds.  As a
consequence, one gets again $P(t,r) = 1 - S_0(t,r)$.

\subsection*{Other results}

\subsubsection*{Mean first-binding time}

The mean first-binding time reads
\begin{align}
& \langle \t \rangle_r = \tilde{S}_0(0,r) 
 = \frac{2r \rho(R^3 - \rho^3)/\mu + 2\rho R^3(r-\rho) - r \rho^2(r^2-\rho^2)}{6r\rho^2 D_0} , 
 \label{eq:MFPT_M0}
\end{align}
while the mean excursion time (at $r = \rho$) is
\begin{equation}
\label{eq:Tmean}
\langle \t \rangle_{\rho} = \frac{R^3 - \rho^3}{3D_0 \rho \mu} = \frac{\rho V}{\mu D_0 A} \,.
\end{equation}
where $V$ is the volume of the domain $\Omega_0$ and $A$ is the area
of the sensor.  As a consequence, the mean first-binding time, which
is essentially proportional to the volume of the active zone, is a
useless characteristics in this situation.  In turn, the mode (i.e.,
the position of the density maximum, i.e., the most probable value) is
representative.

\subsubsection*{Asymptotic analysis of the smallest eigenvalue}

The long-time behavior of $\psi_1(t,r)$, $Q(t)$, and the probability
$P(t,r)$, is determined by the smallest absolute value of the pole
$|p_1|$ of the underlying Laplace-transformed quantity.  Let us first
consider the density $\psi_1(t,r)$, for which the smallest $|p_1|$ is
determined by $\hat{\alpha}_1$.  Denoting $x = \hat{\alpha}_1 \beta$ and
assuming that $x\to 0$, one can use the Taylor expansion of
Eq. (\ref{eq:alpha0_M0}) to determine the asymptotic behavior of
$\hat{\alpha}_1$ for large $\beta$.  In the lowest order in $1/\beta$, we
get
\begin{equation}
\hat{\alpha}_1^2 \simeq \frac{3\mu}{(1+\mu)\beta^3} \simeq \frac{3\mu \rho^3}{(1+\mu) R^3} \,.
\end{equation}
According to Eq. (\ref{eq:A_S_final}), the above relation determines
the slowest decay rate of the survival probability, $\rho^2/(D_0
\hat{\alpha}_1^2)$, which is close to the mean time (\ref{eq:Tmean})
when $\rho \ll R$.

\subsubsection*{Short-time asymptotic behavior}

The short-time asymptotic behavior corresponds to the limit $p\to
\infty$.  In this limit, Eq. (\ref{eq:tildeP_final}) becomes in the
leading order in $1/p$:
\begin{equation}  \label{eq:tildeP0_asympt}
\tilde{P}(p,r) \simeq \frac{\mu \sqrt{D_0} \, \exp\bigl(-(r-\rho)\sqrt{p/D_0}\bigr)}{r \, p^{3/2}}  ,
\end{equation}
from which the short-time asymptotic behavior follows for $r > \rho$
\begin{equation}
P(t,r) \simeq \frac{4(D_0t)^{3/2} \mu}{\sqrt{\pi} r (r-\rho)^2}\, \exp\bigl(-(r-\rho)^2/(4D_0t)\bigr)  .
\end{equation}
This asymptotic behavior is applicable at times as short as $t \ll
(r-\rho)^2/(4D_0)$.  
In turn, for $r = \rho$, Eq. (\ref{eq:tildeP0_asympt}) yields
\begin{equation}
P(t,\rho) \simeq \frac{2\sqrt{D_0} \mu}{\sqrt{\pi} \rho} \, t^{1/2}  \quad (t\to 0)\,.
\end{equation}

\section{One buffer case}
\label{sec:AexampleM1}

For a single buffer ($M = 1$), Eq. (\ref{eq:F_delta}) reads
\begin{equation} \label{eq:F_oneM}
H = \bigl(p + k_{01} - (D_0/\rho^2) z\bigr)\bigl(p + k_{10} - (D_1/\rho^2) z\bigr) - k_{01} k_{10} ,
\end{equation}
and its two zeros determine $\delta_0$ and $\delta_1$:
\begin{align}
\label{eq:delta_1M}
\delta^2_0 & = \frac{\rho^2}{2D_0 D_1} \biggl(D_0(p+k_{10})+D_1(p+k_{01}) 
 - \sqrt{(D_0(p+k_{10})-D_1(p+k_{01}))^2 + 4D_0D_1 k_{01} k_{10}}\biggr) , \\
\delta^2_1 & = \frac{\rho^2}{2D_0 D_1} \biggl(D_0(p+k_{10})+D_1(p+k_{01}) 
 + \sqrt{(D_0(p+k_{10})-D_1(p+k_{01}))^2 + 4D_0D_1 k_{01} k_{10}}\biggr) . 
\end{align}
Getting 
\begin{equation}
\label{eq:fp_M1}
\begin{split}
f(p) & = C_{00} C_{11} - C_{01} C_{10},  \\
f_{00}(p) & = C_{11}, \quad f_{01}(p) = -C_{10}  \\
\end{split}
\end{equation}
from the $2\times 2$ matrix $C$, one obtains the coefficients $b_{0j}$
\begin{equation}
\label{eq:b00b01}
\begin{split}
b_{00} & = \frac{C_{11}\, \mu }{p(C_{00}C_{11} - C_{01} C_{10})} \,, \\
b_{01} & =  - \frac{C_{10} \, \mu}{p(C_{00}C_{11} - C_{01} C_{10})} \,, \\
\end{split}
\end{equation}
where the elements $C_{ij}$ are given explicitly by Eq. (\ref{eq:C}).
We obtain thus
\begin{equation}
\label{eq:tildeS0_M1}
\tilde{S}_0(p,r) = \frac{1}{p} + b_{00} \, v(\delta_0,r) + b_{01} \, v(\delta_1,r) .
\end{equation}

In order to invert the Laplace transform, one needs to determine the
poles of $\tilde{S}_0(p,r)$ that are given by the zeros $\hat{p}_n$ of
the function $f(p)$.  There are infinitely many zeros and they are
nonpositive: $\hat{p}_n \leq 0$.  To compute the residues, one needs
the derivative of $f(p)$ with respect to $p$, which can be evaluated
by using Eq. (\ref{eq:dc_ij}) and
\begin{equation}
\frac{\partial \delta_j}{\partial p} = \frac{\rho^2}{2 \delta_j} \frac{2p+k_{01} +k_{10} - (D_0+D_1)\delta_j^2/\rho^2}
{D_0(p+k_{10})+D_1(p+k_{01}) - 2D_0D_1 \delta_j^2/\rho^2} . 
\end{equation}

Finally, we proceed to check that the two zeros of $f(p)$, $p = 0$ and
$p = -(k_{01} + k_{10})$, are not the poles of $\tilde{S}_0(p,r)$, and
thus excluded from the analysis.

(i) In the limit $p \to 0$, we get
\begin{subequations}  \label{eq:asympt_p0}
\begin{align}
\delta_0^2 & \simeq \rho^2 \frac{k_{01} + k_{10}}{D_0 k_{10} + D_1 k_{01}} p + O(p^2) ,\\
\delta_1^2 & \simeq \frac{\rho^2 (D_0 k_{10} + D_1 k_{01})}{D_0D_1} + O(p), \\
v(\delta_0,\rho) & \simeq - \delta_0 - \delta_0^3 (\beta^3/3 + \beta^2/2) + O(\delta_0^5), \\
\nonumber
C_{00} & \simeq \mu \delta_0 + \delta_0^3 \biggl(\frac{\beta^3}{3}+\beta^2+\beta + \mu\biggl(\frac{\beta^3}{3} 
+\frac{\beta^2}{2}\biggr)\biggr) + O(\delta_0^5) , \\
C_{10} & \simeq \delta_0^3 \biggl(\frac{\beta^3}{3}+\beta^2+\beta\biggr) + O(\delta_0^5), 
\end{align}
\end{subequations}
whereas $C_{01}$ and $C_{11}$ approach constants.  We get thus
\begin{equation}
\begin{split}
b_{00} & \simeq \frac{\mu}{p}\, \frac{C_{11}}{\mu \delta_0 C_{11} - O(\delta_0^3)} = \frac{1}{\delta_0 p} + O(p^{-1/2}), \\
b_{01} & \simeq - \frac{\mu}{p} \, \frac{O(\delta_0^3)}{\mu \delta_0 C_{11} - O(\delta_0^3)} = O(1) . \\
\end{split}
\end{equation}
Since $v(\delta_0,r) = - \delta_0 + O(\delta_0^3)$, the singularities
from $a_0 = 1/p$ and $b_0 v(\delta_0,r)$ cancel each other so that $p
= 0$ is not a pole of $\tilde S_0(p,r)$.

(ii) Setting $p = - (k_{01} + k_{10}) + \ve$, one has
\begin{equation}
\begin{split}
\delta_0^2 & = - \frac{D_0 k_{01} + D_1 k_{10}}{D_0 D_1} + O(\ve),  \\
\delta_1^2 & = \frac{k_{01} + k_{10}}{D_0 k_{01} + D_1 k_{10}} \ve + O(\ve^2) . \\
\end{split}
\end{equation}
As a consequence, we get 
\begin{equation}
C_{01} \simeq \mu \delta_1 = O(\ve^{1/2}) , \qquad C_{11} \simeq \delta_1^3 = O(\ve^{3/2}) ,
\end{equation}
whereas $a_0$, $C_{00}$ and $C_{10}$ approach constants.  We obtain
then
\begin{equation}
\begin{split}
b_{00} & = \frac{\mu}{p} \, \frac{C_{11}}{C_{00}C_{11} - C_{01}C_{10}} = O(1), \\
b_{01} & = - \frac{\mu}{p} \, \frac{C_{10}}{C_{00}C_{11} - C_{01}C_{10}} \simeq \frac{\mu}{C_{01} p} \simeq \frac{1}{\delta_1 p} = O(\ve^{-1/2}). \\
\end{split}
\end{equation}
Since $v(\delta_1,r) \simeq - \delta_1$, the term $b_{01}
v(\delta_1,r)$ has no singularity so that $p = -(k_{01} + k_{10})$ is
not a pole of $\tilde S_0(p,r)$.

We conclude that
\begin{equation}
S_0(t,r) = \sum\limits_{n=1}^\infty \biggl(\hat{b}^0_n \, v\bigl(\delta_0(\hat{p}_n),r\bigr) 
+ \hat{b}^1_n \, v\bigl(\delta_1(\hat{p}_n),r\bigr)\biggr) \exp(\hat{p}_n t) ,
\end{equation}
where $v(\delta,r)$ is given by Eq. (\ref{eq:vr}), and
\begin{equation}
\label{eq:bn0_M1}
\hat{b}^0_n  = \frac{\mu \, C_{11}(\hat{p}_n)}{\hat{p}_n \, f'(\hat{p}_n)} , \qquad
\hat{b}^1_n  = - \frac{\mu  \, C_{10}(\hat{p}_n)}{\hat{p}_n \, f'(\hat{p}_n)} . 
\end{equation}
The derivative with respect to $t$ yields
\begin{align} 
\psi_1(t,r) & = \sum\limits_{n=1}^\infty \biggl(\hat{b}^0_n \, v\bigl(\delta_0(\hat{p}_n),r\bigr) 
+ \hat{b}^1_n \, v\bigl(\delta_1(\hat{p}_n),r\bigr)\biggr)  \times |\hat{p}_n| \exp(\hat{p}_n t) .
\end{align}

\subsection*{Probability $P(t,r)$}

Similarly, the inversion of $\tilde{P}(p,r)$ involves the zeros $p_n$
of $F(p)$ from Eq. (\ref{eq:Fp}) that can be written explicitly as:
\begin{equation}
\label{eq:Fp_M1}
F(p) = (p + \koff) f(p) + \koff \mu \bigl(C_{11} \, v(\delta_0, \rho) - C_{10} \, v(\delta_1,\rho) \bigr),
\end{equation}
with $f(p)$ from Eq. (\ref{eq:fp_M1}).  As previously, one can show
that the zero $p = -(k_{01} + k_{10})$ is not a pole of
$\tilde{P}(p,r)$.  In turn, $p = 0$ is a pole.  In fact, using
Eqs. (\ref{eq:asympt_p0}), one get as $p\to 0$
\begin{equation*}
\begin{split}
F(p) & \simeq \delta_0 p \biggl\{\mu \, C_{11}(0) + \lambda \frac{D_0(k_{01}+k_{10})}{D_0 k_{10} + D_1 k_{01}} 
\biggl(\frac{\beta^3}{3}+\beta^2+\beta\biggr)  \times \bigl(C_{11}(0) - C_{01}(0) - \mu \, v(\delta_1(0),\rho)\bigr) \biggr\}, \\
\end{split}
\end{equation*}
where $C_{01}(0)$, $C_{11}(0)$ and $\delta_1(0)$ denote the values of
these functions evaluated at $p = 0$.  In turn,
\begin{align*}
w(p,r) & = \mu \bigl(C_{11} v(\delta_0,r) - C_{10} v(\delta_1,r)\bigr)  \simeq - \mu C_{11}(0) \delta_0 + O(\delta_0^3) ,
\end{align*}
so that the residue at $p=0$ is
\begin{align} 
P_{\infty} & = \biggl\{1 + \lambda \frac{D_0(k_{01}+k_{10})}{D_0 k_{10} + D_1 k_{01}} 
\, \frac{(1+\beta)^3-1}{3} \times \frac{C_{11}(0) - C_{01}(0) - \mu v(\delta_1(0),\rho)}{\mu C_{11}(0)} \biggr\}^{-1} , 
\end{align}
which is independent of the starting point $r$.  After
simplifications, we have
\begin{equation}   \label{eq:PinfA_M1}
P_{\infty} = \biggl(1 + \lambda \, \frac{(1+\beta)^3-1}{3\mu} (1 + k_{01}/k_{10}) \biggr)^{-1}. 
\end{equation}
This expression coincides with Eq. (\ref{eq:PinfA}).

We get thus
\begin{equation}
P(t,r) = P_\infty + \sum\limits_{n=1}^\infty \biggl(b_n^0 \, v(\delta_0(p_n),r) + b_n^1 \, v(\delta_1(p_n),r)\biggr) e^{p_n t},
\end{equation}
with
\begin{equation}
\label{eq:bn_M1}
b_n^0 = - \frac{\mu C_{11}(p_n)}{F'(p_n)} , \qquad  b_n^1 = \frac{\mu C_{10}(p_n)}{F'(p_n)} .
\end{equation}

\subsection*{One fixed buffer}

For the fixed buffer ($D_1 \to 0$), Eqs. (\ref{eq:delta_1M}) yield
\begin{equation}
\label{eq:asympt_auxil}
\delta^2_0 = \frac{\rho^2}{D_0} \biggl(p + k_{01} - \frac{k_{01} k_{10}}{p + k_{10}}\biggr),  \qquad \delta_1^2 \to \infty .
\end{equation}
As a consequence, one needs to treat this case separately to avoid
diverging terms.

The last relation in Eqs. (\ref{eq:asympt_auxil}) implies that 
\begin{equation*}
c_{i1} \simeq \frac12 (1+\beta) \delta_1^2 e^{\beta \delta_1} \to \infty  \qquad (i=0,1).  
\end{equation*}
In addition, we have
\begin{equation*}
C_{00} = c_{00}, \qquad C_{10} = \frac{c_{10} k_{10}}{p + k_{10}} ,  \qquad C_{11} = \frac{c_{11} k_{10}}{p + k_{10}} ,
\end{equation*}
so that in the limit $D_1 \to 0$, we get
\begin{equation}
b_{00} = \frac{\mu}{p \, c_{00}},  \qquad b_{01} = 0 ,
\end{equation}
given that
\begin{equation}
\frac{C_{01}}{C_{11}} = \frac{c_{01}}{c_{11} \frac{k_{10}}{p + k_{10} - (D_1/\rho^2) \delta_1^2}} = 
\frac{c_{01}}{c_{11} \frac{p + k_{10} - (D_0/\rho^2) \delta_1^2}{k_{01}}} \to 0 .
\end{equation}
We conclude that
\begin{equation}
\label{eq:tildeS_1I}
\tilde{S}_0(p,r) = \frac{1}{p} + \frac{\mu \, v(\delta_0,r)}{p \, c_{00}} ,
\end{equation}
i.e., we retrieved the solution (\ref{eq:tildeS_nobuf}) for the case
without buffer, in which $\delta_0 = \rho\sqrt{p/D_0}$ is replaced by
$\delta_0 = \rho \sqrt{p'/D_0}$, where
\begin{equation}
\label{eq:pprime}
p' = p + k_{01} - \frac{k_{01} k_{10}}{p + k_{10}} .
\end{equation}

The fixed buffer is expected to slow down the arrival onto the sensor
because of binding calcium ions and thus stopping their diffusion.  In
particular, one can notice this effect in an increase of the mean
first-binding time to the sensor, given by $\tilde S_0(0,r)$.  Noting
that $p' = 0$ from Eq. (\ref{eq:pprime}) at $p = 0$, one finds that
the mean first-binding time without buffer, $\langle \t_{nb}\rangle$,
is multiplied by the factor $(1 + k_{01}/k_{10})$ in the presence of a
fixed buffer:
\begin{equation}
\label{eq:MFPT_fixed}
\langle \t \rangle = \tilde{S}_0(0,r) = \langle \t_{nb}\rangle \biggl(1 + \frac{k_{01}}{k_{10}}\biggr).
\end{equation}

The relation to the former solution without buffer allows one to
easily invert the Laplace transform.  In fact, the former poles of
$\tilde S_0(p,r)$ were $\hat{p}_n = - D_0 \hat{\alpha}_n^2/\rho^2$.
Inverting the relation (\ref{eq:pprime}), one can see that each former
pole $\hat{p}_n$ splits in two new poles $\hat{p}_{n,1} =
-\lambda_{n,1}$ and $\hat{p}_{n,2} = -\lambda_{n,2}$, with
\begin{subequations}
\begin{align}
\lambda_{n,1} & = \frac{\sigma_n - \sqrt{\sigma_n^2 - 4k_{10} D_0 \hat{\alpha}_n^2/\rho^2}}{2} \,, \\
\lambda_{n,2} & = \frac{\sigma_n + \sqrt{\sigma_n^2 - 4k_{10} D_0 \hat{\alpha}_n^2/\rho^2}}{2} \,, 
\end{align}
\end{subequations}
with $\sigma_n = D_0 \hat{\alpha}_n^2/\rho^2 + k_{01} + k_{10}$.  As a
consequence, the inverse Laplace transform of Eq. (\ref{eq:tildeS_1I})
becomes
\begin{equation}
\label{eq:S_final_immobile}
S_0(t,r) = \sum\limits_{n=1}^\infty \hat{b}_n \, u(\hat{\alpha}_n,r) \biggl(c_{n,1} \, e^{- \lambda_{n,1} t} + c_{n,2} \, e^{- \lambda_{n,2} t} \biggr) , 
\end{equation}
where $u(\delta,r)$ is given by Eq. (\ref{eq:urA}), the coefficients
$\hat{b}_n$ are given by Eq. (\ref{eq:bn0_M0}), and the weights
\begin{subequations}
\begin{align}
c_{n,1} & = \frac{D_0 \hat{\alpha}_n^2/\rho^2}{\lambda_{n,1}}\, \frac{1}{1 + \frac{k_{01}k_{10}}{(\lambda_{n,1}-k_{10})^2}} \,, \\ 
c_{n,2} & = \frac{D_0 \hat{\alpha}_n^2/\rho^2}{\lambda_{n,2}}\, \frac{1}{1 + \frac{k_{01}k_{10}}{(\lambda_{n,2}-k_{10})^2}} \,, 
\end{align}
\end{subequations}
appear from the change of variables: $dp'/dp = 1 + k_{01} k_{10}/(p +
k_{10})^2$, see Eq. (\ref{eq:pprime}), and from the factor $1/p$ in
the second term of Eq. (\ref{eq:tildeS_1I}).  Note that if $k_{01} =
k_{10} = 0$, one has $\lambda_{n,1} = 0$ and $\lambda_{n,2} = D_0
\hat{\alpha}_n^2/\rho^2$, and one retrieves Eq. (\ref{eq:A_S_final}).

Note also that $\lambda_{n,1} \to k_{10}$ and $\lambda_{n,2} \to D_0
\hat{\alpha}_n^2/\rho^2$ as $n\to \infty$ and thus $c_{n,1} \to 0$ and
$c_{n,2} \to 1$.  In other words, the exchange kinetics does not
affect the high-frequency eigenmodes.

Substituting Eq. (\ref{eq:S_final_immobile}) into (\ref{eq:tildePPA},
\ref{eq:tildeQ}), we get
\begin{equation}
\tilde{P}(p,r) = - \frac{\mu \, v(\delta_0,r)}{(p+\koff) c_{00} + \koff \, \mu \, v(\delta_0,\rho)} ,
\end{equation}
so that one needs to find zeros of the denominator of this expression.
As in the former case for $\tilde{S}_0(p,r)$, one can expect two
sequences of zeros: $p_{n,1} \to - k_{10}$ and $p_{n,2} \to -\infty$.
In fact, when $p \to -k_{10} + 0$, $p'$ from Eq. (\ref{eq:pprime})
diverges to $-\infty$, so that there are infinitely many zeros
accumulating towards $-k_{10}$.  This accumulation requires a more
subtle numerical procedure to calculate zeros.


\section{Calcium channel model used for Monte Carlo simulations}

We describe a VGCC by a 3-state Hodgkin and Huxley gating model
\cite{hodgkin1952quantitative} so that the calcium release was modeled
according to:
\begin{equation}
C_0 \mathrel{\mathop{\rightleftarrows}^{2\alpha{(V(t))}}_{\beta(V(t))}} C_1 
\mathrel{\mathop{\rightleftarrows}^{\alpha{(V(t))}}_{2 \beta(V(t))}} O
\mathrel{\mathop{\rightarrow}^{k(t)}} Ca^{2+}  ,
\end{equation}
with two closed states $C_0$, $C_1$ and one open state $O$ of the
VGCC.  Here $\alpha(V(t))$ and $\beta(V(t))$ are voltage dependent
rates, computed as
\begin{equation}  
\begin{array}{l} 
\alpha(V(t))= \exp(V(t)/20.5) , \\ \beta(V(t))= 0.14 \, \exp(-V(t)/15), \\  \end{array} 
\end{equation}
for a given AP waveform $V(t)$ in mV.  The dynamics starts from the
close state $C_0$.  The parameters in these rates were adjusted such
that the resulting single channel open probability, current duration,
and peak match experimentally observed quantities
\cite{nakamura2015nanoscale}.

The calcium ions are released from the open channel with the rate:
\begin{equation}
k(t)= \frac{g}{2e} (V(t)- V_{\rm rev}) ,
\end{equation}
where $g = 3.3$~pS (picoSiemen) is the single channel conductance
\cite{Sheng12}, $e$ is the elementary charge, and $V_{\rm rev} =
-45$~mV is the reversal potential \cite{nakamura2015nanoscale}.

%

\newpage
\section{Supplementary Figures}

\renewcommand{\figurename}{Supplementary Figure}

\begin{figure*}[!th]
\centering
\includegraphics[width=0.9\linewidth]{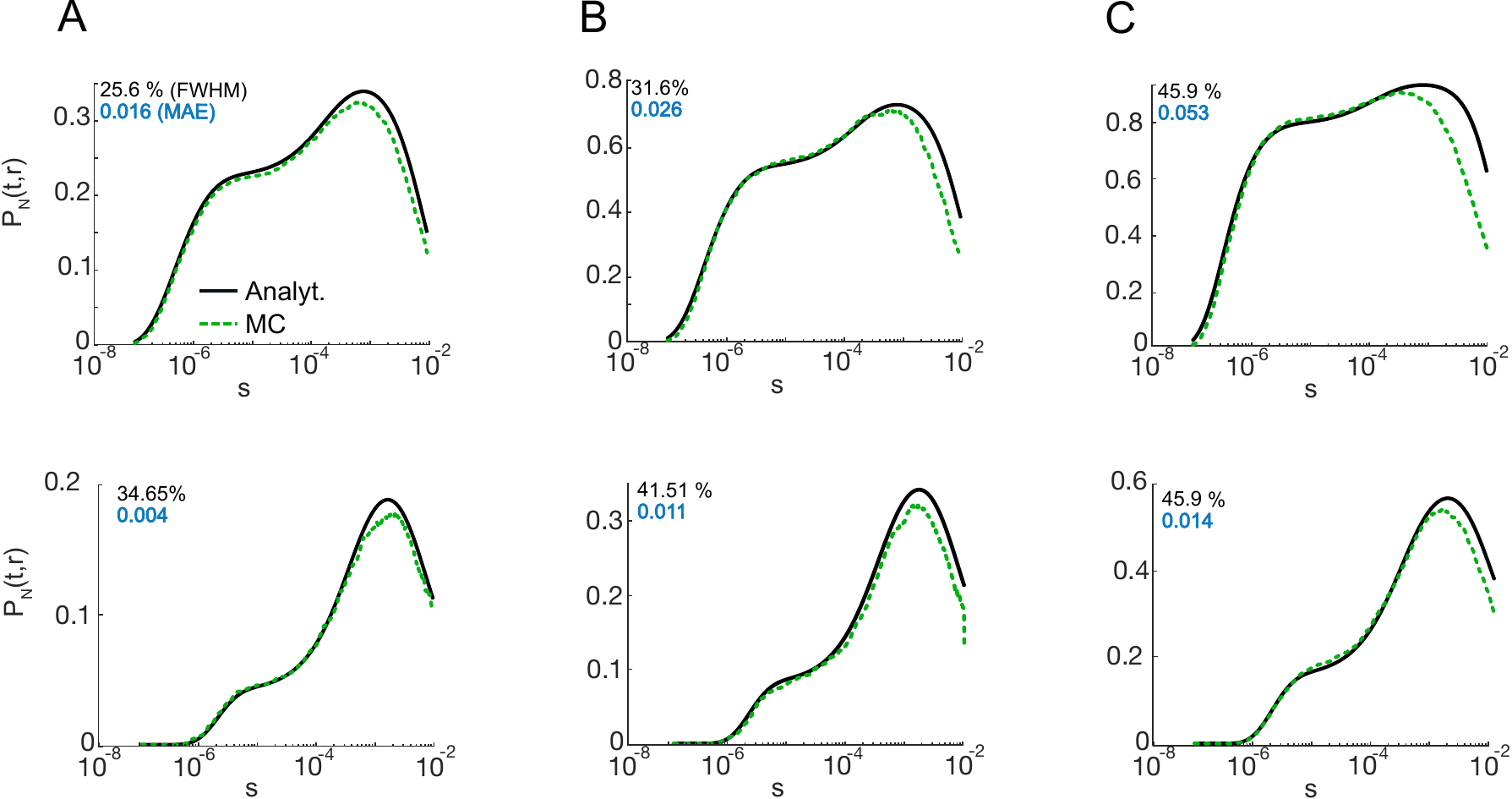}
\caption{\label{fig:Sup1} $P(t,r)$ with slow unbinding kinetics  ($\kon=0.157~\konU$) for 50 (\textbf{A}), 100 (\textbf{B}) and 200 (\textbf{C}) simultaneously released ions for CD of 15 nm (top row) and 45 nm (bottom row). Black and green lines show respectively analytical and MC results. The black and blue inset text on each plot represent FWHM error and MAE between analytical and MC results correspondingly. }
\end{figure*}

\begin{figure*}[!th]
\centering
\includegraphics[width=0.9\linewidth]{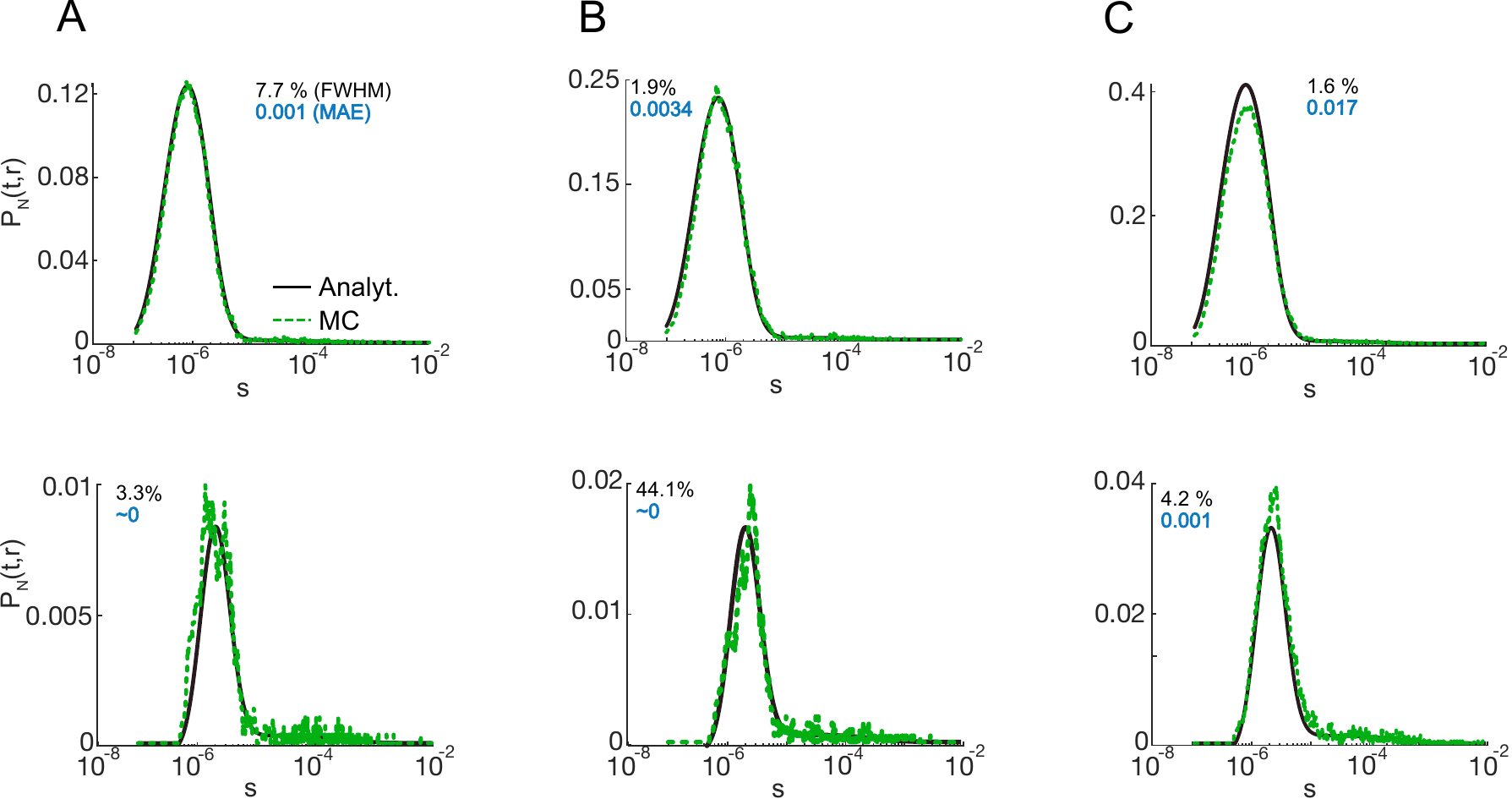}
\caption{\label{fig:Sup2} $P(t,r)$ with fast unbinding kinetics  ($\kon=1570~\konU$) for 50 (\textbf{A}), 100 (\textbf{B}) and 200 (\textbf{C}) simultaneously released ions for CD of 15 nm (top row) and 45 nm (bottom row). Black and green lines show respectively analytical and MC results. The black and blue inset text on each plot represent FWHM error and MAE between analytical and MC results correspondingly. }
\end{figure*}

\end{document}